\documentclass[onecolumn,aps,prc,superscriptaddress]{revtex4}
\usepackage{amssymb}
\usepackage{amsmath}
\usepackage{graphicx}
\usepackage[normalem]{ulem}
\usepackage{multirow}
\usepackage{appendix}
\usepackage[usenames]{color}
\usepackage{bm}
\usepackage[colorlinks,linkcolor=blue,citecolor=blue,urlcolor=blue]{hyperref}
\begin{document}
\title{Electromagnetic fields from the extended Kharzeev-McLerran-Warringa model\\ in relativistic heavy-ion collisions}
\author{Yi Chen}
\affiliation{Shanghai Institute of Applied Physics, Chinese Academy of Sciences, Shanghai 201800, China}
\affiliation{University of Chinese Academy of Sciences, Beijing 100049, China}
\author{Xin-Li Sheng}
\affiliation{Key Laboratory of Quark and Lepton Physics (MOE) and Institute of Particle Physics,\\
Central China Normal University, Wuhan 430079, China}
\author{Guo-Liang Ma}\email[Electronic address: ]{glma@fudan.edu.cn}
\affiliation{Key Laboratory of Nuclear Physics and Ion-beam Application (MOE),Institute of Modern Physics,
Fudan University, Shanghai 200433, China}
\affiliation{Shanghai Institute of Applied Physics, Chinese Academy of Sciences, Shanghai 201800, China}
\begin{abstract}

        Based on the Kharzeev-McLerran-Warringa (KMW) model that estimates strong electromagnetic (EM) fields generated in relativistic heavy-ion collisions, we generalize the formulas of EM fields in the vacuum by incorporating the longitudinal position dependence, the generalized charge distributions and retardation correction. We further generalize the formulas of EM fields in the pure quark-gluon plasma (QGP) medium by incorporating a constant Ohm electric conductivity and also during the realistic early-time stages QGP evolution by using a time-dependent electric conductivity. Using the extended KMW model, we observe a slower time evolution and a more reasonable impact parameter $b$ dependence of the magnetic field strength than those from the original KMW model in the vacuum. The inclusion of medium effects by using the lattice data helps to further prolong the time evolution of magnetic field, such that the magnetic field strength during the realistic QGP evolution at thermal freeze-out time can meet the $1\sigma$ bound constrained from experimentally measured difference in global polarizations of $\Lambda$ and $\bar{\Lambda}$ hyperons in Au+Au collisions at top RHIC energy. These generalized formulations in the extended KMW model will be potentially useful for many EM fields relevant studies in heavy-ion collisions, especially at lower colliding energies and for various species of colliding nuclei.

\end{abstract}

\maketitle

\section{Introduction}

        The study of strongly interacting matter and its properties in the presence of strong electromagnetic (EM) fields has been a hot topic for more than a decade~\cite{Rafski1976,Kharzv2008,Skokov2009,Tuchin2010,Sergei2010,Askawa2010,Voroyk2011,LiOou2011,Bzdak2012,DenHu2012,ToneA2012,ToneB2012,TuchiC2013,TuchiR2013,Blocz2013,
        Ypeng2014,Gursoy2014,McLerr2014,Zahaov2014,Tuchin2015,DenHu2015,Blocz2015,VecRoy2015,LiHui2016,Rober2017,Balth2017,DuanSh2018,Gursoy2018,XLZhao2018,XLZhaB2019,XLZhaC2019,YChen2019,Hamelm2019,XianG2020,BangX2020}. Ultra-relativistic heavy-ion collisions provide a unique way for creating and exploring the strongly interacting matter at extremely high temperature and high energy density, where the matter is expected to be deconfined and reach a new state of matter, which is so-called the ``quark-gluon plasma" (QGP)~\cite{Shuryk1980,HwaBK2004,Adams2005,Bzdak2019}. Properties of strongly interacting matter are governed by quantum chromodynamics (QCD), which have been widely and experimentally studied both on the Relativistic Heavy Ion Collider (RHIC) at Brookhaven National Laboratory (BNL) and on the Large Hadron Collider (LHC) at CERN. In heavy-ion collisions, e.g., at top RHIC energy at $\sqrt{s}=200\mathrm{\,GeV}$ or LHC energy at $\sqrt{s}=5.02\mathrm{\,TeV}$, the two oppositely fast moving (almost at the speed of light) colliding nuclei in non-central nucleus-nucleus (A-A) collisions can generate hitherto the strongest EM fields~\cite{Rafski1976,Kharzv2008,Skokov2009,Tuchin2010,Sergei2010,Askawa2010,Voroyk2011,LiOou2011,Bzdak2012,DenHu2012,ToneA2012,ToneB2012,TuchiC2013,TuchiR2013,Blocz2013,
        Ypeng2014,Gursoy2014,McLerr2014,Zahaov2014,Tuchin2015,DenHu2015,Blocz2015,VecRoy2015,LiHui2016,Rober2017,Balth2017,DuanSh2018,Gursoy2018,XLZhao2018,XLZhaB2019,XLZhaC2019,YChen2019,Hamelm2019,XianG2020,BangX2020}, which are usually estimated to be at the order of magnitude of $eB \sim eE \sim m_{\pi}^2 \sim 10^{18}\mathrm{\,G}$ at top RHIC energy, or $eB \sim eE \sim 10\,m_{\pi}^2 \sim 10^{19}\mathrm{\,G}$ at LHC energies, where $m_{\pi}$ is the pion mass. Here we should note that event-by-event fluctuations of generated EM fields have been widely studied in recent years in Refs.~\cite{Bzdak2012,DenHu2012,ToneA2012,Blocz2013,DenHu2015,Blocz2015,VecRoy2015,XLZhao2018,XLZhaB2019,XLZhaC2019,Hamelm2019,YChen2019,XianG2020}, which can give rise to non-vanishing components of EM fields such as $|B_{x}|$ and $|E_x|$ due to the fluctuations of proton positions in the two colliding nuclei.

        In the QCD vacuum, topologically non-trivial gluon field configurations with non-zero winding number $Q_{\mathrm{w}}$~\cite{Belavn1975,tHooft1975,Jackiw1976} of deconfined QGP in the presence of such a strong magnetic field $\bf{B}$ can induce a non-conserved axial current $j^{\mu}_5=\sum_{f}q_f\langle \bar{\psi}_{f}\gamma^{\mu}\gamma^{5}\psi_{f} \rangle$ and also a non-conserved vector current $j^{\mu}= \sum_{f}q_f\langle \bar{\psi}_{f}\gamma^{\mu}\psi_{f} \rangle$ along the direction of magnetic field, which are respectively called the ``chiral separation effect" (CSE) and ``chiral magnetic effect" (CME)~\cite{Kharzv2008,Askawa2010,Fukush2008,Warrig2008,Kharzv2009,FukusA2010,FukusL2010,Kharzv2010,Kharzv2013,Kharzv2014,Koichi2017}. Since the axial current $j^{\mu}_5$ requires a charge-conjugate $\mathcal{C}$-odd environment while vector current $j^{\mu}$ requires a chirality imbalanced parity $\mathcal{P}$-odd environment, an asymmetry between the amount of positive and negative charges along the direction of magnetic field $\bf{B}$ in heavy-ion collisions is expected. Experimental observations of the CME can be regarded as direct evidences of topologically non-trivial gluon field configurations, and furthermore can be interpreted as indications of event-by-event local $\mathcal{P}$ and $\mathcal{CP}$ violations of QCD at quantum level~\cite{Kharzv2008,Fukush2008}.

        Besides the CME and CSE, it is well known that strong magnetic fields can also influence many QCD processes~\cite{Kharzv2013}, e.g. the induction of chiral symmetry breaking~\cite{Gusyni1994}, influences on chiral condensation~\cite{Mizher2010}, and modifications of in-medium particle's mass~\cite{HaoLei2016,Bonati2017,GSBali2018,Coppol2018,Shijun2019,Ferrer2019,HTDing2020}. As an important consequence, the QCD phase diagram may be dynamically modified by such a strong magnetic field~\cite{Anders2016,Stefa2018,Masim2018,GShao2019,Shile2020}, e.g. color-superconducting phases at very high baryon densities will be strongly affected by the magnetic field~\cite{Anders2016}. When anomaly processes are coupled with strong magnetic fields, many interesting effects~\cite{Kharzv2013}, e.g. the formation of $\pi^0$-domain walls~\cite{DTSon2008}, will also be induced and generated.

        Since the generated magnetic field $\bf{B}$ in heavy-ion collisions can not be directly measured event-by-event, it is thus of enormous challenge to measure the magnetic field induced chiral anomalous effects in experiments. In heavy-ion collisions, the magnetic field is generated along the direction roughly but preferentially perpendicular to the reaction plane (RP), experimental measurements are therefore usually conducted using the two-particle correlator $\gamma_{\alpha\beta}=\langle\langle{\cos(\phi_{\alpha}+\phi_{\beta}-2\Psi_{\mathrm{RP}}}\rangle\rangle$ firstly proposed by Voloshin~\cite{Volsh2004}, where $\alpha$ and $\beta$ denote the electric charge sign of particles $\alpha$ and $\beta$, $\phi_{\alpha}$ and $\phi_{\beta}$ are respectively their azimuthal angles, $\Psi_{\mathrm{RP}}$ is the azimuthal angle of the constructed reaction plane for a given event, and $\langle\langle \cdots \rangle\rangle$ denotes the average over all particle pairs and then over all events. Therefore, the same-sign (SS) and opposite-sign (OS) correlators can be respectively defined as $\gamma_{\mathrm{ss}} \equiv (\gamma_{++}+\gamma_{--})/2$ and $\gamma_{\mathrm{os}} \equiv (\gamma_{+-}+\gamma_{-+})/2$. Based on Refs.~\cite{Sergei2010,Volsh2004}, the magnetic field driven CME is expected to contribute to a negative $\gamma_{\mathrm{ss}}$ but a positive $\gamma_{\mathrm{os}}$. The STAR Collaboration~\cite{Selyu2006,STARL2009,STARC2010,Volsh2011,STARA2013,STARL2014} and ALICE Collaboration~\cite{ALICE2014} have independently measured the $\gamma_{\mathrm{ss}}$ and $\gamma_{\mathrm{os}}$ correlators, which indeed show the expected features of the CME. However, there exit some ambiguities~\cite{FWang2010,Bzdak2010,JLiao2010,Bzdak2011,Schli2010,SchlP2010,Schli2011,Pratt2011,Bzdak2013} in the interpretation of experimental data due to large background contaminations, potentially arising from the elliptic-flow ($v_2$) driven background contributions, e.g., the transverse momentum conservation (TMC)~\cite{Bzdak2011,Pratt2011,Bzdak2013} and local charge conservation (LCC)~\cite{Schli2010,SchlP2010,Schli2011}. Hence, a dedicated run of $\mathrm{^{96}Zr+^{96}Zr}$ and $\mathrm{^{96}Ru+^{96}Ru}$ isobar colliding systems at RHIC has been proposed~\cite{Sergei2010}, which is expected to yield unambiguous evidence for the CME signal by varying the signal but with the $v_2$-driven backgrounds roughly fixed~\cite{Sergei2010,WDeng2017,VKoch2017,Huang2017,SZShi2018,haoXu2018,SZShi2019}.

        In this paper, we will review an analytical model for the estimation of strong EM fields generated in relativistic heavy-ion collisions from the original work initiated by Kharzeev, McLerran and Warringa~\cite{Kharzv2008}, which we refer to as the original KMW model. On the ground of it, we formulate our generalizations of the estimated EM fields. We first start from generalizing the three-dimensional charge distributions, e.g. three-parameter Fermi (3pF) model, by incorporating Lorentz contraction effect, based on which we point out that the formulas of EM fields in the original KMW model can be properly extended by incorporating the longitudinal position dependence through the generalized charge distribution models for both spherically symmetric and axially deformed colliding nuclei used in heavy-ion collisions. Also, we make retardation correction to the estimated EM fields contributed by participants. Thus our formulation of the estimated EM fields can be easily applied to lower energy regions, such as the current beam-energy-scan (BES) program at RHIC, the under planning FAIR, NICA and J-PARC programs. It is due to the fact that the Lorentz contraction is not so large that the ``pancake-shaped disk" approximation used in the original KMW model~\cite{Kharzv2008} is no longer appropriate in these lower energy regions. Moreover, we further extend the formulas of EM fields by incorporating medium feedback effects according to the Faraday's induction law, thus a constant Ohm conductivity $\sigma_{0}$ is properly and analytically embedded for the pure QGP medium and also a time-dependent conductivity $\tilde{\sigma}(t)$ for the realistic QGP evolution. Finally, we make numerical evaluations of the generalized magnetic field strength in the extended KMW model for detailed comparisons of time evolution, impact parameter $b$ dependence of the estimated magnetic field strength with those from the original KMW model.

        This paper is organized as follows. We present detailed formulations of the estimated EM fields in the extended KMW model for heavy-ion collisions in Sec. \ref{sec:Electromagnetic Fields from the Extended KMW Model}, which consists of four subparts. We first present in Sec. \ref{sec:Generalization of Charge Distributions for Heavy-ion Collisions} a formal generalization of the charge distributions from the widely used 3pF model in which the relativistic Lorentz contraction effects on the geometries of both spherically formed and axially deformed colliding nuclei are taken into account. We then in Sec. \ref{sec:Electromagnetic Fields from the Extended KMW Model in the Vacuum} generalize the EM fields in the vacuum starting from the widely used Li$\acute{\mathrm{e}}$nard-Wiechert equations, and naturally extend the formulas of EM fields with generalized charge distributions and retardation correction. We further extend the EM fields with medium feedback effects by incorporating a constant Ohm electric conductivity $\sigma_{0}$ in Sec. \ref{sec:Electromagnetic Fields from the Extended KMW Model in Pure Conducting Medium}, a time-dependent electric conductivity in Sec. \ref{sec:Electromagnetic Fields from the Extended KMW Model during QGP Evolution} and also an alternative solution for simulations in \ref{sec:An Alternative Solution to the Estimated EM Fields in Simulations}. Some evaluations and comparisons about the time evolution, centrality (impact parameter $b$) dependence of the estimated magnetic field from the extended KMW model in comparison with those from the original KMW model are presented in Sec.~\ref{sec:Results and Discussions}. We finally summarize the main process of such generalization along with the conclusions in Sec.~\ref{sec:Summary}. The notation we use in this paper is the rationalized Lorentz-Heaviside units within the natural units, with $\hbar=c=1$ and $\mu_0=\epsilon_0=1$.

\section{Electromagnetic Fields from the Extended KMW Model}
\label{sec:Electromagnetic Fields from the Extended KMW Model}

        Before the detailed discussion, let us first emphasize that physical situations of heavy-ion collisions consist of event-by-event local $\mathcal{P}$ and $\mathcal{CP}$ violation processes due to the effects of topological charge fluctuations with non-trivial QCD gauge field configurations (characterized by the topological invariant, the winding number $Q_{\mathrm{w}}$) in the vicinity of the deconfined QGP phase~\cite{Kharzv2008}. These $\mathcal{P}$ and $\mathcal{CP}$ violation processes can only locally happen under some special and even extreme conditions in QCD, such as in the instantaneous deconfined QGP phase with chirality imbalance at extremely high temperature $T \sim \Lambda_{\mathrm{QCD}} $ in the presence of a strong magnetic field $eB \sim m_{\pi}^2 $, which will equivalently generate event-by-event locally non-vanishing $\theta$ angle for the effects of so called ``$\theta$-vacuum"~\cite{Khazv2006,Khazv2007,Calan1976}. In other words, fluctuations of local $\mathcal{P}$ and $\mathcal{CP}$ violated metastable domains of the QCD $\theta$ vacuum on the event-by-event basis are intrinsically connected to the amount of charges $Q_{e}$ separated by the simultaneous magnetic field $\bf{B}$ for the non-trivial gluon field configurations with $Q_{\mathrm{w}} \neq 0$. It means that the expectation value of the amount of separated charges $\langle Q_{e} \rangle$ is proportional to the strength of magnetic field $|\mathbf{B}|$, hence the charge asymmetry fluctuations $\langle \Delta_{\pm}^{2} \rangle$ will be roughly proportional to $|\mathbf{B}|^2$~\cite{Kharzv2008}. It is therefore of crucial importance to quantify the strength of EM fields, especially the time evolution and the possible lifetime $t_{\mathrm{B}}$ of the magnetic field $\mathbf{B}(t, \mathbf{r})$ in relativistic heavy-ion collisions~\cite{Stewa2018,Berndt2018}.

\begin{figure}[htb]
            \includegraphics[angle=0,scale=0.72]{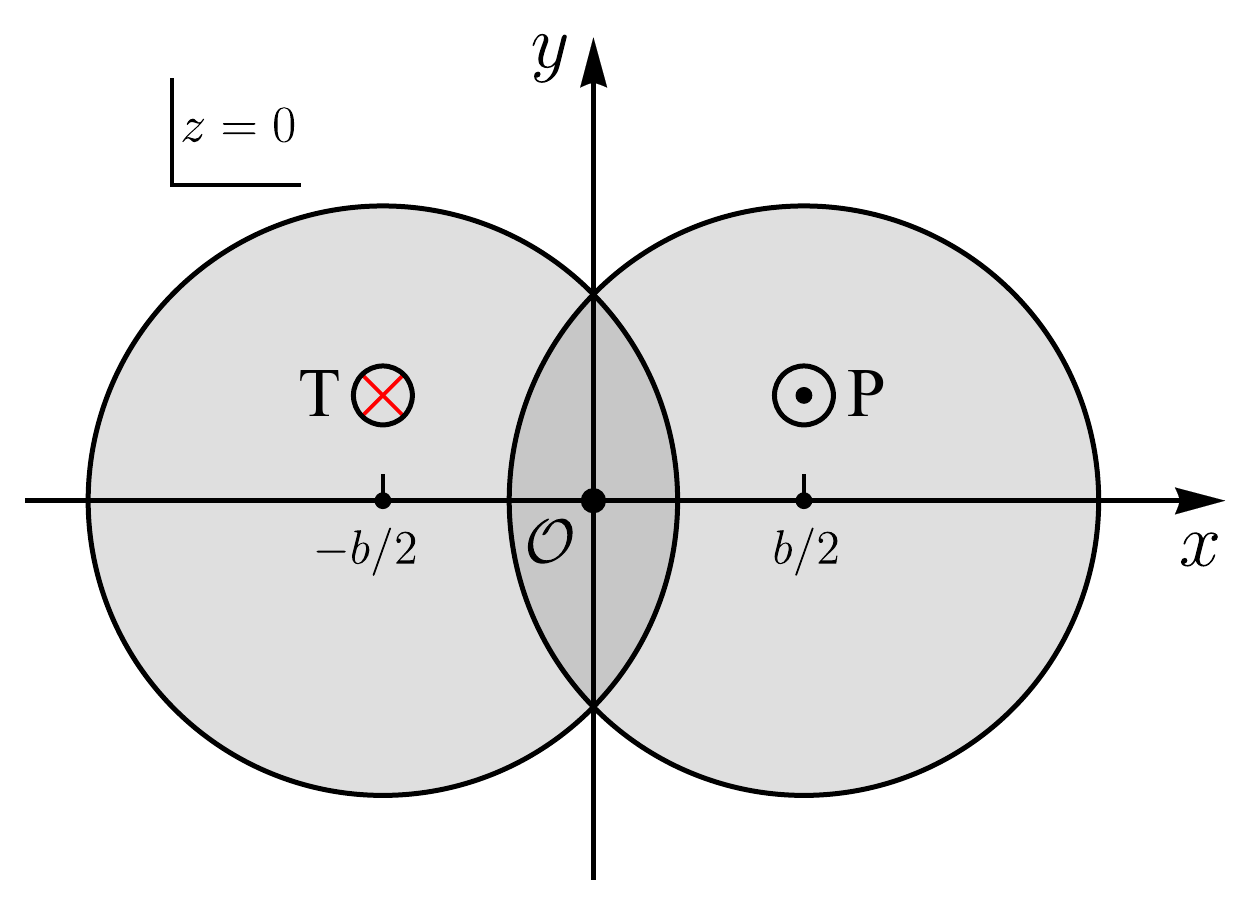}
            \caption{(Color online) Illustration of the initial geometry of the colliding system projected on the transverse plane ($z=0$) at an impact parameter $b$ for non-central high-energy nucleus-nucleus collisions. The centers of the projectile (denoted as P) and target (denoted as T) nucleus are respectively located at $(b/2, 0, 0)$ and $(-b/2, 0, 0)$ at $t=0$ when the two colliding nuclei are completely overlapping with each other, moving parallel or anti-parallel to the beam $z$ direction.}
            \label{Fig01_CS}
        \end{figure}

        Similar to the conventional setup of colliding system for heavy-ion collisions, we choose the $x$ axis along the impact parameter $b$, and $z$ axis along the beam direction of projectile such that the $x-z$ plane is exactly the reaction plane. The $y$ axis is then chosen to be perpendicular to the reaction plane, as illustrated in Fig. \ref{Fig01_CS}.

\subsection{Generalization of Charge Distributions for Heavy-ion Collisions}
\label{sec:Generalization of Charge Distributions for Heavy-ion Collisions}

        Since in symmetric heavy-ion collisions such as gold-gold collisions at top RHIC energy with the center of mass energy $\sqrt{s}=200\mathrm{\,GeV}$ per nucleon pair along the $z$ direction, the Lorentz contraction factor $\gamma=\sqrt{s}/(2m_{\mathrm{N}}) \simeq 106.5$ (where $m_{\mathrm{N}}$ is the average nucleon mass), which corresponds to the beam rapidity $Y_0=\cosh^{-1}(\gamma) \simeq 5.36$. In this case, the two gold nuclei will be Lorentz contracted to be less than $1\%$ of their original size in the $z$ direction, because of which the original KMW model~\cite{Kharzv2008} approximates the two colliding nuclei as ``pancake-shape disk". It is further assumed in this model that the charge distribution can be treated as uniformly distributed within the two colliding nuclei and further be limited only within the transverse plane with a two-dimensional surface number density. Therefore in this paper, we will first formulate our generalization of the charge distributions, and then systematically formulate the estimation of generated EM fields in the extended KMW model.

        It has been widely acknowledged that final state anisotropies of emitted charged particles in heavy-ion collisions are very sensitive to the initial conditions, such as the experimentally measured final state elliptic flow $v_2$ of charged particles is actually very sensitive to the initial eccentricity $\varepsilon_{2}$~\cite{QYShou2018}. Hence, the initial conditions may dominantly cause the contamination of the experimentally measured CME signal. Therefore, a better quantification of the initial geometry will be of crucial importance, especially for the study of CME in heavy-ion collisions since the initial geometry of charge distributions may significantly modify not only the centrality dependence but also the time evolution of the generated magnetic field $\mathbf{B}(t, \mathbf{r})$.

        Besides, due to the relativistic motion of colliding nuclei, the overall size of each nucleus will be Lorentz contracted by a factor of $\gamma$ along the beam $z$ direction, such that a spherically symmetric nucleus will become an ellipsoidal nucleus with the volume shrunken by a factor of $\gamma$ while the charge density magnified by a factor of $\gamma$ simultaneously. Hence, we first start from generalizing the charge distributions into a Lorentz contracted three-dimensional form with the corresponding standard parameters obtained from high-energy electron-scattering measurements~\cite{DVries1987}. Moreover, what the original high-energy electron-scattering experiments measure is actually the charge spatial distribution in the so-called ``Breit Frame"~\cite{Ernst1960} rather that the nuclear density profile with relativistic motion. Hence, we note that the originally inferred charge distribution parameters from electron-scattering experiments should therefore be properly modified~\cite{QYShou2018} so as to be unambiguously implanted into the widely used Monte-Carlo simulations of the initial conditions for the finite-number nucleons sampling of each nucleus~\cite{Miller2007}.

        For non-central high-energy nucleus-nucleus (A-A) collisions at center of mass energy $\sqrt{s}=2\gamma m_{\mathrm{N}}$ per nucleon pair with the same setup of colliding system as illustrated in Fig. \ref{Fig01_CS}, the charge distribution used in the original KMW model can be generalized from the widely used charge distribution models listed in~\cite{DVries1987}, e.g. the widely accepted and used three-parameter Fermi model (3pF) for heavy ions, with the relativistic Lorentz contraction effect taken into account. For a source point $\mathbf{r}^{\,\prime}=(x^{\prime},y^{\prime},z^{\prime})=(\mathbf{r}^{\,\prime}_{\perp},z^{\prime})$ within a colliding nucleus (with charge $Ze$, and $e=|e|$) with its center located at $\mathbf{r}_c^{\,\prime}=(0, 0, 0)$, the generalized charge number density reads~\cite{DuanSh2018,DVries1987},

        \begin{equation}\label{3pF_01}
        \begin{aligned}
        \rho(\mathbf{r}^{\,\prime}) = \gamma\rho_0\frac{1+\omega\left[r^{\prime}/R^{\prime}(\theta,\phi)\right]^2}{1+\exp \big{\{}[r^{\prime}-R^{\prime}(\theta,\phi)]/d\big{\}}}\Theta(\mathbf{r}^{\,\prime}),
        \end{aligned}
        \end{equation}
        where
        \begin{equation}\label{3pF_02}
        \begin{aligned}
        r^{\prime} = r^{\prime}(\gamma) \equiv \sqrt{ {\mathbf{r}^{\,\prime}_{\perp}}^2 + (\gamma z^{\prime})^2 }.
        \end{aligned}
        \end{equation}

        In the expression of $\rho(\mathbf{r}^{\,\prime})$ in Eq. (\ref{3pF_01}), the polar and azimuthal angle variations of axially deformed $R^{\prime}(\theta,\phi)$ is defined as \cite{Blocz2015,WDeng2017,Huang2017}
        \begin{equation}\label{3pF_03}
        \begin{aligned}
        R^{\prime}(\theta,\phi)=R_0\left[1+\beta_2 Y_2^0\left(\theta\right)+\beta_4 Y_4^0\left(\theta\right) + \cdots \right],
        \end{aligned}
        \end{equation}
        where $R_0$ is the spherical monopole charge radius, and $Y_{l}^{m}(\theta,\phi)$ denotes the spherical harmonics, with $\theta$ being the polar angle with respect to the symmetry axis of each nucleus and $\phi$ being the corresponding azimuthal angle, which are respectively defined as $\phi\equiv \sin^{-1}(y^{\prime}/|\mathbf{r}^{\,\prime}_{\perp}|)$ and $\theta \equiv \cos^{-1}(z^{\prime}/r^{\prime})$. The diffusion depth of the nuclear surface $d$ can be roughly related to the skin thickness $t$ by $t=4\ln(3)d\simeq 4.4\,d$, where $t$ equals the distance in which the charge density falls from $90\%$ to $10\%$ of its central value $\rho_0$.

        Note that $\omega$ in Eq. (\ref{3pF_01}) is an additional parameter~\cite{DVries1987}, introduced to characterize the charge central depression or elevation phenomena of some nuclide such as $\mathrm{^{40}Ca}$, relative to the traditional two-parameter Fermi model (2pF) with parameter $c$ (which is in analogue to $R^{\prime}$ in 3pF model) and $d$. Only in 3pF model with $\omega=0$ that the half density radius $R_{1/2}=R^{\prime}(\theta,\phi)$, which is the distance from the center of the nucleus to the point at which the charge density equals half its central value $\rho_0$. $\rho_0$ is usually regarded as the charge number normalization constant, which can be obtained from the following charge number normalization condition
        \begin{equation}\label{3pF_04}
        \begin{aligned}
        Z=\int_{-{R_{\mathrm{A}}}}^{+R_{\mathrm{A}}} \mathrm{d}x^{\prime} \int_{-{R_{\mathrm{A}}}}^{+R_{\mathrm{A}}} \mathrm{d}y^{\prime} \int_{-{R_{\mathrm{A}}/\gamma}}^{+R_{\mathrm{A}}/\gamma} \mathrm{d}z^{\prime}\,  \rho(\mathbf{r}^{\,\prime}),
        \end{aligned}
        \end{equation}
        where $R_{\mathrm{A}}$ is the radius of each nucleus. One should note that the $\gamma$ factor in front of $\rho_0$ in Eq. (\ref{3pF_01}) is due to the fact that since the volume of each colliding nucleus will be Lorentz contracted by a factor of $\gamma$ in the $z$ direction, in accordance to which the charge number density will be simultaneously enlarged by a factor of $\gamma$ at the same time.

        The function $\Theta(\mathbf{r}^{\,\prime})$ is introduced so as to constrain that the charge density is just within the ellipsoidal colliding nucleus, and it can further be used to separate spectators from participants when restricting it just within the transverse plane, namely $\Theta(\mathbf{r}^{\,\prime}) \to \tilde{\Theta}(\mathbf{r}_{\perp}^{\,\prime})$, which are separately defined as
        \begin{equation}\label{3pF_06}
        \begin{aligned}
        \Theta(\mathbf{r}^{\,\prime}) &\equiv \theta \left[ R_{\mathrm{A}}^2 - {\mathbf{r}^{\,\prime}_{\perp}}^2 - (\gamma z^{\prime})^2  \right],\\
        \tilde{\Theta}(\mathbf{r}_{\perp}^{\,\prime}) &\equiv \theta \left[ R_{\mathrm{A}}^2 - {\mathbf{r}^{\,\prime}_{\perp}}^2 \right].
        \end{aligned}
        \end{equation}
        Here we note that $\Theta(\mathbf{r}^{\,\prime})$ and $\tilde{\Theta}(\mathbf{r}_{\perp}^{\,\prime})$ are similar to the function $\theta(\mathbf{r}_{\perp}^{\,\prime})$ introduced in the original KMW model~\cite{Kharzv2008}. The function $\theta(x)$ in Eq. (\ref{3pF_06}) is the standard unit step function, which reads
        \begin{equation}\label{3pF_08}
        \begin{aligned}
        \theta(x)=
        \begin{cases}
        0,& x<0,\\
        1,& x>0.
        \end{cases}
        \end{aligned}
        \end{equation}

        For a spherically symmetric nucleus ($\beta_2=\beta_4=0$), e.g. the $\mathrm{^{197}Au}$ nucleus, $R^{\prime}$ is actually independent of polar and azimuthal angle distributions and is just the monopole charge radius, namely $R^{\prime}=R_0$ (similar to the parameter $c$ in 2pF model), as a consequence of which charge number density $\rho(\mathbf{r}^{\,\prime})=\rho({r}^{\prime})$ is only a function of ${r}^{\prime}(\gamma)$. In the relativistic situation under this condition, the relativistic three-dimensional density $\rho(\mathbf{r}^{\,\prime})$ according to Eq. (\ref{3pF_01}) reads
        \begin{equation}\label{3pF_09}
        \begin{aligned}
        \rho(\mathbf{r}^{\,\prime}) = \gamma\rho_{0}\frac{1+\omega(r^{\prime}/R_0)^2}{1+\exp \big{[}\left( r^{\prime}-R_0 \right)/d\big{]}}\Theta(\mathbf{r}^{\,\prime}).
        \end{aligned}
        \end{equation}
        However, for an axially deformed nucleus, e.g. the $\mathrm{^{96}Zr}$ nucleus, its shape may deviate from the ideal sphere and the charge distribution will depend on the polar and azimuthal angle distributions, for which Eq. (\ref{3pF_03}) is to some extent introduced so as to account for such kind of charge distribution inside the 3pF model as in Eq. (\ref{3pF_01}). In the characterization of deformations in Eq. (\ref{3pF_03}), $\beta_2 Y_2^0$ and $\beta_4 Y_4^0$ characterize angular variations of $R^{\prime}(\theta,\phi)$ of quadrupole and hexadecapole deformations respectively, where $\beta_2$ and $\beta_4$ are the corresponding quadrupole and hexadecapole deformation parameters, and $Y_2^0(\theta)$ and $Y_4^0(\theta)$ are the corresponding spherical harmonics.

\subsection{Electromagnetic Fields from the Extended KMW Model in the Vacuum}
\label{sec:Electromagnetic Fields from the Extended KMW Model in the Vacuum}

        Since we have generalized the charge number density in a relativistic three-dimensional form in Eqs. (\ref{3pF_01}) or (\ref{3pF_09}), let us now first quote a simple situation where EM fields are generated by a pair of two constantly but oppositely moving point-like charged particles with same electric charge $Ze$ (where $e$ is the elementary electric charge, $e=|e|$) but different rapidity $Y_{n}=\pm Y$ ($Y>0$), which corresponds to the velocity $\mathbf{v}_{n}=(0, 0, \tanh Y_{n})$ parallel or anti-parallel to the $z$ direction. We now set that at $t=0$ the two point-like charged particles are located at $\mathbf{r}^{\,\prime}_{n}(t=0)=(x^{\prime}_{n}, y^{\prime}_{n}, z^{\prime}_{n})$. Here the index $n=\pm$ respectively represents the charged particles moving in the positive and negative $z$-directions. Also, it is worth to note that since in the three-dimensional situation in heavy-ion collisions for a charge point $\mathbf{r}^{\,\prime}_{n}$ within the two colliding nuclei at time $t=0$ when the two colliding nuclei are usually assumed to be completely overlapping with each other, the $z$ coordinate $z^{\prime}_{n}$ for this charge point in general will be non-zero. One can start from using the Li$\acute{\mathrm{e}}$nard-Wiechert (L-W) potentials generated by two point-like charge particles to evaluate the EM fields for a field point $\mathbf{r}=(x, y, z)$ at observation time $t$, which gives rise to the following widely used L-W equations of EM fields as~\cite{Skokov2009,Sergei2010,LiOou2011,Bzdak2012,DenHu2012,ToneA2012,Blocz2013,Ypeng2014,DenHu2015,Blocz2015,VecRoy2015,XLZhao2018,XLZhaB2019,XLZhaC2019,Hamelm2019,YChen2019,XianG2020}
        \begin{equation}\label{LS_BS01}
        \begin{aligned}
        e \mathbf{E}(t, \mathbf{r}) &= \alpha_{\mathrm{EM}} \sum_{n=\pm} Z_{n} \frac{\mathbf{R}_{n}\left(1-v_{n}^{2}\right)}{\left(R_{n}^2-\left[\mathbf{v}_{n} \times \mathbf{R}_{n}\right]^2\right)^{3/2}},\\
        e \mathbf{B}(t, \mathbf{r}) &= \alpha_{\mathrm{EM}} \sum_{n=\pm} Z_{n}  \frac{\mathbf{v}_{n} \times \mathbf{R}_{n}\left(1-v_{n}^{2}\right)}{\left(R_{n}^2-\left[\mathbf{v}_{n} \times \mathbf{R}_{n}\right]^2\right)^{3/2}},
        \end{aligned}
        \end{equation}
        where $\mathbf{R}_{n}=\mathbf{r}-\mathbf{r}^{\,\prime}_{n}(t)$ is the relative position vector from a field point $\mathbf{r}$ to a source point $\mathbf{r}^{\,\prime}_{n}$ at the same observation time $t$, $\mathbf{r}^{\,\prime}_{n}(t)=(x^{\prime}_{n}, y^{\prime}_{n}, z^{\prime}_{n})+\mathbf{v}_{n}t$ is the source position of a point-like charge at observation time $t$, and $\alpha_{\mathrm{EM}}$ is the EM fine-structure constant, defined as $\alpha_{\mathrm{EM}}\equiv e^2/4\pi \approx 1/137$. One should note that the retardation effect due to field propagation has already been incorporated into above equations. For later use convenience, above Eq. (\ref{LS_BS01}) can be directly rewritten, in terms of rapidity $Y$, in the following form,
        \begin{equation}\label{LS_BS02}
        \begin{aligned}
        e \mathbf{E}(t, \mathbf{r}) &= Z\alpha_{\mathrm{EM}} \sum_{n=\pm} { \frac{ \cosh(Y_n) \cdot \mathbf{R}_{n} }{ \Big{\{} (x-x^{\prime}_{n})^2 + (y-y^{\prime}_{n})^2 + \left[ t\sinh Y_n - (z-z^{\prime}_{n})\cosh Y_n \right]^2 \Big{\}}^{3/2} } },\\
        e \mathbf{B}(t, \mathbf{r}) &= Z\alpha_{\mathrm{EM}} \sum_{n=\pm} {   \frac{\sinh(Y_n) \cdot \mathbf{e}_{z} \times \mathbf{R}_{n} }{ \Big{\{} (x-x^{\prime}_{n})^2 + (y-y^{\prime}_{n})^2 + \left[ t\sinh Y_n - (z-z^{\prime}_{n})\cosh Y_n \right]^2 \Big{\}}^{3/2} } },
        \end{aligned}
        \end{equation}
        which in nature are well consistent with the derivation in~\cite{Kharzv2008} for the magnetic field generated by a point-like charge when reducing $z^{\prime}_{n}\to 0$ in above Eq. (\ref{LS_BS02}). In contrary, $z^{\prime}_{n} \neq 0$ is intended for the three-dimensional form of charge distribution such as Eqs. (\ref{3pF_01}) or (\ref{3pF_09}) where the $z$ coordinate of a point-like charge inside a colliding nuclei is not necessarily limited only within the transverse plane, which thus helps to generalize the ``pancake shaped disk" approximation as well as the two-dimensional surface number density used in the original KMW model~\cite{Kharzv2008}.

        Now one can estimate the strength of EM fields generated in heavy-ion collisions. For a field point $\mathbf{r}=(x,y,z)=(\mathbf{r}_{\perp}, z)$ at observation time $t$, the EM fields generated by two identical colliding nuclei with charge $Ze$ and beam rapidity $\pm Y_0$ ($Y_0>0$) parallel or anti-parallel to the $z$ direction can be estimated by combining the generalized relativistic three-dimensional charge number density $\rho(\mathbf{r}^{\,\prime})$ in Eqs. (\ref{3pF_01}) or  (\ref{3pF_09}) with the EM fields generated by a pair of two point-like charges in Eq. (\ref{LS_BS02}), and then performing the integration of charge number densities over the corresponding coordinate space within the two colliding nuclei. Nevertheless, the situation will be a little different from the point-like charge case since colliding two heavy ions will separate the charged bulk matter into spectators (denote as S) and participants (denote as P). Participants in general will slow-down to a certain extent due to the baryon-junction stopping effect~\cite{Kharzv1996}, while spectators in general are supposed to keep moving with the same beam rapidity $Y_{b}=\pm Y_0$ and do not scatter at all. Therefore, we follow the method proposed in~\cite{Kharzv2008} by likewise splitting the contributions of EM fields generated by two colliding nuclei into four parts, which simply read
        \begin{equation}\label{LS_BS05}
        \begin{aligned}
        \mathbf{E} &= \mathbf{E}_{\mathrm{S}}^{+} + \mathbf{E}_{\mathrm{S}}^{-} + \mathbf{E}_{\mathrm{P}}^{+} + \mathbf{E}_{\mathrm{P}}^{-},\\
        \mathbf{B} &= \mathbf{B}_{\mathrm{S}}^{+} + \mathbf{B}_{\mathrm{S}}^{-} + \mathbf{B}_{\mathrm{P}}^{+} + \mathbf{B}_{\mathrm{P}}^{-}.
        \end{aligned}
        \end{equation}

        For a charge position vector $\mathbf{r}^{\prime}=(x^{\prime},y^{\prime},z^{\prime})=(\mathbf{r}_{\perp}^{\,\prime},z^{\prime})$ with the same setup of colliding system as in Fig. \ref{Fig01_CS}, the charge number density $\rho(\mathbf{r}^{\,\prime})$ in Eqs. (\ref{3pF_01}) or (\ref{3pF_09}) as well as the corresponding $\Theta(\mathbf{r}^{\,\prime})$ and $\tilde{\Theta}(\mathbf{r}_{\perp}^{\,\prime})$ functions should be accordingly shifted by a half impact parameter vector $\mathbf{b}$, namely $\rho_{\pm}(\mathbf{r}^{\,\prime})=\rho(\mathbf{r}^{\,\prime}\mp\mathbf{b}/2)$, $\Theta_{\pm}(\mathbf{r}^{\,\prime}) = \Theta(\mathbf{r}^{\,\prime}\mp\mathbf{b}/2)$ and $\tilde{\Theta}_{\pm}(\mathbf{r}_{\perp}^{\,\prime})=\tilde{\Theta}(\mathbf{r}_{\perp}^{\,\prime}\mp\mathbf{b}/2)$. The contributions of EM fields from spectators can therefore be elaborated as
        \begin{equation}\label{LS_SP01}
        \begin{aligned}
        e\mathbf{E}_{\mathrm{S}}^{\pm}(t, \mathbf{r}) &=  \alpha_{\mathrm{EM}}\cosh(Y_0)\int_{V_{\pm}} \mathrm{d}^3\mathbf{r}^{\,\prime} \frac{ \rho_{\pm}(\mathbf{r}^{\,\prime})\left[1- \tilde{\Theta}_{\mp}(\mathbf{r}_{\perp}^{\,\prime})\right] \mathbf{R}_{\pm} }{ \Big{\{} (\mathbf{r}_{\perp}-\mathbf{r}_{\perp}^{\,\prime})^2  + \left[ t\sinh (\pm Y_0) - (z-z^{\prime})\cosh Y_0 \right]^2 \Big{\}}^{3/2} },\\
        e\mathbf{B}_{\mathrm{S}}^{\pm}(t, \mathbf{r}) &=  \alpha_{\mathrm{EM}}\sinh(\pm Y_0)\int_{V_{\pm}} \mathrm{d}^3\mathbf{r}^{\,\prime} \frac{ \rho_{\pm}(\mathbf{r}^{\,\prime})\left[1- \tilde{\Theta}_{\mp}(\mathbf{r}_{\perp}^{\,\prime})\right] \mathbf{e}_{z} \times \mathbf{R}_{\pm} }{ \Big{\{} (\mathbf{r}_{\perp}-\mathbf{r}_{\perp}^{\,\prime})^2  + \left[ t\sinh (\pm Y_0) - (z-z^{\prime})\cosh Y_0 \right]^2 \Big{\}}^{3/2} },
        \end{aligned}
        \end{equation}
        where $V_{\pm}$ donate the ellipsoidal volumes occupied by the two colliding nuclei, which can be related to the integration domains for charge number normalization conditions as in Eq. (\ref{3pF_04}).

        For EM fields contributed by participants, we at present do not consider contributions of newly created particles either as Ref.~\cite{Kharzv2008} because the numbers of newly produced positively and negatively charged particles will be nearly equal and their expansion is roughly spherical. Such an assumption should be reasonable, especially for peripheral collisions. Therefore one only needs to take into account the contributions of baryon-junction stopping bulk matter which are initially there~\cite{Kharzv2008}. The normalized rapidity distributions of the baryon-junction stopping bulk matter of projectile ($+$) and target ($-$) can be empirically estimated as~\cite{Kharzv2008,Gursoy2014,Gursoy2018,Kharzv1996}
        \begin{equation}\label{LS_PT01}
        \begin{aligned}
        f_{\pm}(Y) = \frac{\alpha_{y}}{2\sinh(\alpha_{y} Y_0)} e^{\pm \alpha_{y} Y},\ \ \ \ -Y_0\leq Y\leq Y_0.
        \end{aligned}
        \end{equation}
        Here we note that the parameter $\alpha_{y}$ can be estimated as $\alpha_{y}\simeq \alpha^0_{\mathrm{J}}$, where $\alpha^0_{\mathrm{J}}$ is the Regge trajectory intercept, defined as $\alpha^0_{\mathrm{J}}(0) \simeq 2\alpha_{\mathrm{B}}(0) - 1 + 3\left(1+\alpha_{\mathrm{R}}(0)\right) \simeq 1/2$ in Regge theory~\cite{GRossi1977,GRossi1980,Kharzv1996}, in which $\alpha_{\mathrm{B}}(0)\simeq 0$ is the baryon intercept while $\alpha_{\mathrm{R}}(0)\simeq 0.5$ is the Reggeon intercept. Although it has been mentioned early in~\cite{GRossi1980} that a somewhat lower baryon intercept $\alpha_{\mathrm{B}}(0)$ should be used if nucleon exchange effectively dominates, which will somehow give rise to a relatively smaller $\alpha_{y}$. Recently, the ALICE Collaboration has reported the data at different center of mass energies for the anti-baryons to baryons ratio on baryon stopping~\cite{ALICE2013}, which seems to support that $\alpha_{y}\simeq 0.5$ is experimentally reasonable. Therefore we subjectively to use $\alpha_{y}\simeq 0.5$ as the principal value in this paper, and also check that varying $\alpha_{y}$ between 0.48-0.50 does not largely affect the final results for the time evolution of the magnetic field strength.

        The contributions of EM fields from participants can therefore be estimated by inserting Eq. (\ref{LS_PT01}) into Eq. (\ref{LS_BS02}), and then combining with Eqs. (\ref{3pF_01}) or (\ref{3pF_09}), which follow as
        \begin{equation}\label{LS_PT02}
        \begin{aligned}
        e\mathbf{E}_{\mathrm{P}}^{\pm}(t, \mathbf{r}) &=  \alpha_{\mathrm{EM}}\int_{V_{\pm}} \mathrm{d}^3\mathbf{r}^{\,\prime} \int_{-Y_0}^{Y_0} \mathrm{d}Y \frac{ \Psi_{\pm}(Y)\cosh{Y} \cdot \rho_{\pm}(\mathbf{r}^{\,\prime})\tilde{\Theta}_{\mp}(\mathbf{r}_{\perp}^{\,\prime}) \mathbf{R}_{\pm} }{ \Big{\{} (\mathbf{r}_{\perp}-\mathbf{r}_{\perp}^{\,\prime})^2  + \left[ t\sinh Y - (z-z^{\prime})\cosh Y \right]^2 \Big{\}}^{3/2} },\\
        e\mathbf{B}_{\mathrm{P}}^{\pm}(t, \mathbf{r}) &= \alpha_{\mathrm{EM}}\int_{V_{\pm}} \mathrm{d}^3\mathbf{r}^{\,\prime} \int_{-Y_0}^{Y_0} \mathrm{d}Y \frac{ \Psi_{\pm}(Y)\sinh{Y} \cdot \rho_{\pm}(\mathbf{r}^{\,\prime})\tilde{\Theta}_{\mp}(\mathbf{r}_{\perp}^{\,\prime}) \mathbf{e}_{z} \times \mathbf{R}_{\pm} }{ \Big{\{} (\mathbf{r}_{\perp}-\mathbf{r}_{\perp}^{\,\prime})^2  + \left[ t\sinh Y - (z-z^{\prime})\cosh Y \right]^2 \Big{\}}^{3/2} },
        \end{aligned}
        \end{equation}
        where the refined functions $\Psi_{\pm}(Y)$ for baryon-junction stopping effect are introduced to account for whether the retardation time $t_{\mathrm{ret}}$ of a participating charge at observation time $t$ is ahead of the collision time $t_c$ or simply after the collision time $t_c$, thus functions $\Psi_{\pm}(Y)$ read
        \begin{equation}\label{LS_PT03}
        \begin{aligned}
        \Psi_{\pm}(Y) &= \theta(t_{\mathrm{ret}}-t_c)f_{\pm}(Y) + \theta(t_c-t_{\mathrm{ret}})\delta(Y\mp Y_0).
        \end{aligned}
        \end{equation}
        Here $t_{\mathrm{ret}}$ is obtained by solving the retardation relation $t_{\mathrm{ret}} = t-\big{|}\mathbf{r}-\mathbf{r}^{\prime}-t_{\mathrm{ret}}\cdot \mathbf{e}_{z} \tanh Y \big{|}$ for each rapidity $Y$ within the normalized rapidity integral. We refer to this kind of treatment of participants in Eq. (\ref{LS_PT03}) as retardation correction (RC) in this paper. For the case without retardation correction, the functions $\Psi_{\pm}(Y)$ are simply previously used normalized rapidity distribution functions $f_{\pm}(Y)$ in~\cite{Kharzv2008,Gursoy2014,Gursoy2018}, namely $\Psi_{\pm}(Y)=f_{\pm}(Y)$.

        The expressions for the estimated EM fields in Eqs. (\ref{LS_SP01}) and (\ref{LS_PT02}) are in general consistent with the derivation in the original KMW model~\cite{Kharzv2008}. One distinct difference is that the depth in the $z$ direction is now taken into account in the extended KMW model rather than the infinitely thin depth (zero $z$-depth) in the original KMW model, which actually extends the dimensions of the charge density integration. The other distinct difference is that the retardation correction from the perspective of field propagation for the treatment of participating charges as in Eq. (\ref{LS_PT03}) through $\Psi_{\pm}(Y)$ rather than $f_{\pm}(Y)$ in Refs.~\cite{Kharzv2008,Gursoy2014,Gursoy2018} is now explicitly considered. Hence, we think that our generalization in extended KMW model may give rise to some observable differences for the total estimated EM fields in the vacuum since the retardation effect due to field propagation is most relevant to the $z$ coordinate. Again, we note that the charge number densities $\rho_{\pm}(\mathbf{r}^{\,\prime})$ are not longer the uniformly distributed surface densities, which are now the generalized three-dimensional 3pF model incorporated with Lorentz contraction effect and the effect due to deformations using standard parameters obtained from high-energy electron-scattering measurements~\cite{DVries1987}. Therefore, Eqs. (\ref{LS_SP01}) and (\ref{LS_PT02}) will be very easily and appropriately applied to lower energy regions and various colliding systems, where the ``pancake-shaped disk" approximation used in the original KMW model~\cite{Kharzv2008} is no longer appropriate.

\subsection{Electromagnetic Fields from the Extended KMW Model in Pure Conducting Medium}
\label{sec:Electromagnetic Fields from the Extended KMW Model in Pure Conducting Medium}

        Up to now, we limit our discussions on the estimated EM fields without considering any medium feedback effects, namely the Ohm electric conductivity $\sigma=\sigma_{\mathrm{Ohm}}$ is vanishing in above discussions for the extended KMW model. For high-temperature ($T\sim \Lambda_{\mathrm{QCD}}$) hot QCD matter, the real QGP is a conducting medium and the electric conductivity $\sigma$ is generally acknowledged to be proportional to the plasma temperature $T$~\cite{Arnol2003,Gupta2004,GAart2007,HDing2011,Franc2013,Aless2013,Brand2013}, namely $\sigma\propto T$. Besides, it is known according to the Faraday's induction law that the electric conductivity $\sigma$ may to some extent substantially slow down the decrease of the generated EM fields, and therefore largely prolong the lifetime of EM fields, which is verified to be crucial for the estimation of the CME in heavy-ion collisions~\cite{Stewa2018,Berndt2018}. In order to incorporate the medium feedback effects, we therefore implant a constant electric conductivity $\sigma_{0}$ into the following Maxwell's equations, which read~\cite{LiHui2016}
        \begin{equation}\label{Max_Eq01}
        \begin{aligned}
        \nabla \cdot  \mathbf{E} &=\frac{\rho_{\mathrm{ext}}}{\varepsilon},\\
        \nabla \cdot  \mathbf{B} &=0,\\
        \nabla \times \mathbf{E} &=-\partial_{t}\mathbf{B},\\
        \nabla \times \mathbf{B} &=\partial_{t}\mathbf{E}+\sigma_{0}\mathbf{E}+\mathbf{j}_{\mathrm{ext}}.
        \end{aligned}
        \end{equation}
        We then obtain the following partial differential wave equations for EM fields,
        \begin{equation}\label{Max_Eq02}
        \begin{aligned}
        \left(\nabla^2 - \partial_{t}^2 - \sigma_{0}\partial_{t} \right)\mathbf{B} &= -\nabla \times \mathbf{j}_{\mathrm{ext}},\\
        \left(\nabla^2 - \partial_{t}^2 - \sigma_{0}\partial_{t} \right)\mathbf{E} &= \partial_{t}\mathbf{j}_{\mathrm{ext}} + \nabla\left(\frac{\rho_{\mathrm{ext}}}{\varepsilon}\right).
        \end{aligned}
        \end{equation}
        The solutions for above EM wave equations have been analytically solved in~\cite{LiHui2016}, in which the authors have embedded both electric conductivity $\sigma_{0}$ and chiral magnetic conductivity $\sigma_{\chi}$ for the EM fields generated by a relativistic point-like charge. In the absence of chiral magnetic conductivity $\sigma_{\chi}$ (since in general, $\sigma_{\chi} \ll \sigma_{0}$), we obtain the following solution in the cylindrical coordinates for the magnetic field generated at a field point $\mathbf{r}=(\mathbf{r}_{\perp}, z)=(x, y, z)$ as
        \begin{equation}\label{Max_Eq03}
        \begin{aligned}
        \begin{pmatrix} B_{r}\\ B_{\phi}\\ B_{z}\end{pmatrix} (t,\mathbf{r}) &= \frac{Q}{4\pi}\frac{\gamma v R_{\perp}}{\Delta^{3/2}}\left[1+\frac{\gamma |v|\sigma_{0}}{2}\sqrt{\Delta} \right]e^A \begin{pmatrix} 0\\ 1\\ 0 \end{pmatrix},
        \end{aligned}
        \end{equation}
        with
        \begin{equation}\label{Max_Eq04}
        \begin{aligned}
        R_{\perp} &\equiv |\mathbf{r}_{\perp}-\mathbf{r}_{\perp}^{\,\prime}|=\sqrt{ (x-x^{\,\prime})^2 + (y-y^{\,\prime})^2 },\\
        \Delta    &\equiv \gamma^2(vt+z^{\prime}_0-z)^2+ R_{\perp}^2,\\
        A         &\equiv \frac{\gamma v\sigma_{0}}{2}\left[\gamma(vt+z^{\prime}_0-z)\right] - \frac{\gamma |v|\sigma_{0}}{2}\sqrt{\Delta},
        \end{aligned}
        \end{equation}
        for a point-like charged particle with electric charge $Q=+Ze$ moving with velocity $\mathbf{v}=(0, 0, v)$ along the positive or negative $z$ directions, which is located at $\mathbf{r}^{\,\prime}=(\mathbf{r}_{\perp}^{\prime}, z^{\prime})=(x^{\prime}, y^{\prime}, z^{\prime}_0 + vt)$ at observation time $t$. The radial and longitudinal components of magnetic field $B_{r}$ and $B_{z}$ are both vanishing when only the electric conductivity $\sigma_{0}$ is embedded, namely $B_{r}=B_{z}=0$.

        The transformations of EM fields $\mathbf{F}$ (where $\mathbf{F}=\mathbf{B},\ \mathbf{E}$) in the Cartesian coordinates $(F_x, F_y, F_z)$ with that in the cylindrical coordinates $(F_r, F_{\phi}, F_z)$ simply read
        \begin{equation}\label{Tanfs_Eq01}
        \begin{aligned}
        \mathbf{F}=\begin{pmatrix} F_x\\ F_y\\ F_z \end{pmatrix} &= \begin{pmatrix} F_r\cos\phi-F_{\phi}\sin\phi\\ F_r\sin\phi+F_{\phi}\cos\phi\\ F_z \end{pmatrix},
        \end{aligned}
        \end{equation}
        where $\phi$ is the azimuthal angle (with respect to $x$-axis) of the transverse relative position vector, namely  $\mathbf{R}_{\perp}=\mathbf{r}_{\perp}-\mathbf{r}_{\perp}^{\,\prime}$. Note that since we have $\mathbf{e}_{z} \times \mathbf{R}=(y^{\prime}-y,  x-x^{\prime}, 0)$, where $\mathbf{R}=\mathbf{r}-\mathbf{r}^{\,\prime}$ is the relative position vector,
        the magnetic field $\bf{B}$ in Eq. (\ref{Max_Eq03}) in the Cartesian coordinates therefore can be rewritten in a more compact form, which reads
        \begin{equation}\label{Tanfs_Eq02}
        \begin{aligned}
        \mathbf{B}(t,\mathbf{r}) &= \frac{Q}{4\pi}\frac{\gamma v \mathbf{e}_{z} \times \mathbf{R}}{\Delta^{3/2}}\left[1+\frac{\gamma |v|\sigma_{0}}{2}\sqrt{\Delta} \right]e^A,
        \end{aligned}
        \end{equation}
        which shares almost the same form as the L-W equation of magnetic field $\mathbf{B}(t,\mathbf{r})$ in Eq. (\ref{LS_BS02}) except for the factor $(1+\gamma |v|\sigma_{0}\sqrt{\Delta}/2 )e^A$ due to the inclusion of a constant electric conductivity $\sigma_{0}$ for the conducting medium response. Therefore, Eq. (\ref{Tanfs_Eq02}) can naturally reproduce the well-known L-W equation of magnetic field in the vacuum, i.e., Eq. (\ref{LS_BS01}) or Eq. (\ref{LS_BS02}), when the electric conductivity $\sigma_{0}$ is vanishing. For a special situation in Eq. (\ref{Max_Eq03}) when the azimuthal angle $\phi=0$, the $x$ component of magnetic field $B_x$ is consequently vanishing and the tangential $\phi$ component of magnetic field $B_{\phi}$ is actually $B_{y}$ along the $y$ direction, as is usually mentioned in literature~\cite{TuchiR2013}. Also, we note that the expression of magnetic field $\bf{B}$ in Eq. (\ref{Max_Eq03}) was firstly derived in Ref.~\cite{TuchiR2013,Gursoy2014}.

        For the electric field $\mathbf{E}$, the tangential component of electric field $E_{\phi}$ can be easily obtained by applying the relation $E_{\phi}=-vB_{r}=0$. Meanwhile, the authors in~\cite{LiHui2016} have obtained following explicit expressions for both radial and longitudinal components of electric field $E_r$ and $E_z$ in the leading linear order of $\sigma_{0}$ for a relativistic moving charge as
        \begin{equation}\label{Max_Eq05}
        \begin{aligned}
        E_{r}(t,\mathbf{r}) &= \frac{Q}{4\pi}\left\{\frac{\gamma R_{\perp}}{\Delta^{3/2}}\left(1+\frac{\gamma |v|\sigma_{0}}{2} \sqrt{\Delta}\right) -\frac{\sigma_{0}}{|v|R_{\perp}}\left[1+\frac{\gamma|v|}{\sqrt{\Delta}}\left(t-\frac{z-z^{\prime}_0 }{v}\right) \right]e^{-\sigma_{0}[t-(z-z^{\prime}_0)/v]} \right\}e^A,\\
        E_{\phi}(t,\mathbf{r}) &=0,\\
        E_{z}(t,\mathbf{r}) &= \frac{Q}{4\pi}\left\{ \frac{\sigma_{0}^2 v}{|v|^3} e^{-\sigma_{0}[t-(z-z^{\prime}_0)/v]}\Gamma(0, -A) + \frac{e^A}{{\Delta}^{3/2}}\left[\gamma(z-z^{\prime}_0-vt)-\frac{v}{|v|}A\sqrt{\Delta}-\frac{\gamma v \sigma_{0}}{v^2}\Delta \right]\right\},
        \end{aligned}
        \end{equation}
        where $\Gamma(0, -A) =\int_{-A}^{\infty} \mathrm{d}t \exp(-t)/t $ is the incomplete gamma function. According to the transformations in Eq. (\ref{Tanfs_Eq01}), the radial and longitudinal components of electric field $E_r$ and $E_z$ are respectively along the $x$ and $z$ directions when the azimuthal angle $\phi=0$. Here we note that Eqs. (\ref{Max_Eq03}) and (\ref{Max_Eq05}) can naturally reduce to the well-known L-W equations of EM fields in the vacuum when the electric conductivity $\sigma_{0}$ is vanishing ($\sigma_{0}=0$). It has been further proved in~\cite{LiHui2016} that Eqs. (\ref{Max_Eq03}) and (\ref{Max_Eq05}) are self-consistent and well satisfy the Maxwell's equations in Eq. (\ref{Max_Eq01}). Hence in the following, we will extend above EM fields of a relativistic point-like charge in Eqs. (\ref{Max_Eq03}) and (\ref{Max_Eq05}), and generalize them to be more applicable for high-energy nucleus-nucleus collisions.

        We then combine EM fields in Eqs. (\ref{Max_Eq03}) and (\ref{Max_Eq05}) with the Lorentz contracted three-dimensional charge number density in Eqs. (\ref{3pF_01}) or (\ref{3pF_09}), and then apply the decomposition of EM fields as in Eq. (\ref{LS_BS05}). In analogue to Eqs. (\ref{LS_SP01}) and (\ref{LS_PT02}), we finally obtain the following explicit expressions of the magnetic field in the pure conducting medium. For the contributions of spectators, we obtain
        \begin{equation}\label{Max_SP01}
        \begin{aligned}
        e\mathbf{B}_{\mathrm{S}}^{\pm}(t, \mathbf{r}) &= \lim_{Y\to \pm Y_0} \alpha_{\mathrm{EM}}\sinh Y \int_{V_{\pm}} \mathrm{d}^3\mathbf{r}^{\,\prime} \frac{ \rho_{\pm}(\mathbf{r}^{\,\prime})\left[1- \tilde{\Theta}_{\mp}(\mathbf{r}_{\perp}^{\,\prime})\right] \mathbf{e}_{z} \times \mathbf{R}_{\pm} }{ \Delta^{3/2} }\left[1 + \frac{\sigma_{0}\sinh|Y| }{2}\sqrt{\Delta} \right]e^{A},
        \end{aligned}
        \end{equation}
        with the definitions in Eq. (\ref{Max_Eq04}) accordingly rewritten as
        \begin{equation}\label{Max_SP02}
        \begin{aligned}
        R_{\perp} &\equiv |\mathbf{r}_{\perp}-\mathbf{r}_{\perp}^{\prime}|=\sqrt{ (x-x^{\,\prime})^2 + (y-y^{\,\prime})^2 },\\
        \Delta    &\equiv \left[ t\sinh Y - (z - z^{\prime})\cosh Y \right]^2 + {R_{\perp}}^2,\\
        A         &\equiv \frac{\sigma_{0}\sinh Y }{2}\left[t\sinh Y - (z - z^{\prime})\cosh Y \right] - \frac{\sigma_{0}\sinh|Y|}{2}\sqrt{\Delta},\\
        \end{aligned}
        \end{equation}
        where $\mathbf{R}_{\pm}=\mathbf{r}-\mathbf{r}^{\,\prime}(t)$ are the relative positions of the field point $\mathbf{r}$ to the source point $\mathbf{r}^{\,\prime}$ at the observation time $t$. Again, the positive and negative signs $\pm$ correspond to the nuclei moving in the positive and negative $z$-directions. For the contributions of participants, we have
        \begin{equation}\label{Max_PT01}
        \begin{aligned}
        e\mathbf{B}_{\mathrm{P}}^{\pm}(t, \mathbf{r}) &= \alpha_{\mathrm{EM}}\int_{V_{\pm}} \mathrm{d}^3\mathbf{r}^{\,\prime} \int_{-Y_0}^{Y_0} \mathrm{d}Y \frac{ \Psi_{\pm}(Y)\sinh Y \cdot \rho_{\pm}(\mathbf{r}^{\,\prime})\tilde{\Theta}_{\mp}(\mathbf{r}_{\perp}^{\,\prime}) \mathbf{e}_{z} \times \mathbf{R}_{\pm}  }{ \Delta^{3/2} } \left[1 + \frac{\sigma_{0}\sinh|Y| }{2}\sqrt{\Delta} \right]e^{A}.
        \end{aligned}
        \end{equation}

        For the electric field $\mathbf{E}$ generated by two colliding nuclei, we first notice that $E_{\phi}$ is vanishing in the point-like charge case, thus there is no contribution from both spectators and participants for the tangential component of electric field $E_{\phi}$. The cylindrical coordinates are usually not extensively used in heavy-ion collisions, we therefore directly transform the expressions of electric field to the Cartesian coordinates after invoking Eq. (\ref{Tanfs_Eq01}) as follows,
        \begin{equation}\label{Max_ST01}
        \begin{aligned}
        \begin{pmatrix} e{E}_{x,\mathrm{S}}^{\pm}\\ e{E}_{y,\mathrm{S}}^{\pm} \end{pmatrix}(t, \mathbf{r}) &= \lim_{Y\to \pm Y_0} \alpha_{\mathrm{EM}} \int_{V_{\pm}} \mathrm{d}^3\mathbf{r}^{\,\prime} \rho_{\pm}(\mathbf{r}^{\,\prime})\left[1- \tilde{\Theta}_{\mp}(\mathbf{r}_{\perp}^{\,\prime})\right]\bigg{\{}\frac{ R_{\perp}\cosh Y}{\Delta^{3/2}}\left(1 + \frac{\sigma_{0}\sinh |Y|}{2}\sqrt{\Delta}\right)\\
        \quad\quad\quad&\quad - \frac{\sigma_{0}}{R_{\perp} \tanh |Y|} \left[1 + \frac{\sinh |Y|}{\sqrt{\Delta}} \left(t - \frac{z - z^{\prime}}{\tanh Y}\right) \right] \exp\left(-\sigma_{0}\left[ t - \frac{z-z^{\prime}}{\tanh Y} \right]\right) \bigg{\}}\frac{e^{A}}{R_{\perp}}\begin{pmatrix} x-x^{\prime}\\ y-y^{\prime} \end{pmatrix},\\
        \begin{pmatrix} e{E}_{x,\mathrm{P}}^{\pm}\\ e{E}_{y,\mathrm{P}}^{\pm} \end{pmatrix}(t, \mathbf{r}) &= \alpha_{\mathrm{EM}} \int_{V_{\pm}} \mathrm{d}^3\mathbf{r}^{\,\prime} \int_{-Y_0}^{Y_0} \mathrm{d}Y \ \Psi_{\pm}(Y)\rho_{\pm}(\mathbf{r}^{\,\prime})\tilde{\Theta}_{\mp}(\mathbf{r}_{\perp}^{\,\prime})\bigg{\{}\frac{ R_{\perp}\cosh Y}{\Delta^{3/2}}\left(1 + \frac{\sigma_{0}\sinh |Y|}{2}\sqrt{\Delta}\right)\\
        \quad\quad\quad&\quad - \frac{\sigma_{0}}{R_{\perp} \tanh |Y|} \left[1 + \frac{\sinh |Y|}{\sqrt{\Delta}} \left(t-\frac{z - z^{\prime}}{\tanh Y}\right) \right] \exp\left(-\sigma_{0}\left[ t - \frac{z-z^{\prime}}{\tanh Y} \right]\right)\bigg{\}}\frac{e^{A}}{R_{\perp}}\begin{pmatrix} x-x^{\prime}\\ y-y^{\prime} \end{pmatrix},
        \end{aligned}
        \end{equation}
        which are respectively the contributions of spectators and participants for the $x$ and $y$ components of electric field $E_x$ and $E_y$. For that of the $z$ components of electric field $E_z$, we similarly have
        \begin{equation}\label{Max_ST02}
        \begin{aligned}
        e{E}_{z,\mathrm{S}}^{\pm}(t, \mathbf{r}) &= \lim_{Y\to \pm Y_0} \alpha_{\mathrm{EM}} \int_{V_{\pm}} \mathrm{d}^3\mathbf{r}^{\,\prime} \rho_{\pm}(\mathbf{r}^{\,\prime})\left[1- \tilde{\Theta}_{\mp}(\mathbf{r}_{\perp}^{\,\prime})\right] \bigg{\{}
        \frac{\mathrm{sgn}(Y) \sigma_{0}^2}{\tanh^2 Y} \exp\left(-\sigma_{0}\left[ t - \frac{z-z^{\prime}}{\tanh Y} \right]\right) \Gamma(0, -A)\\
        \quad\quad\quad&\quad + \frac{e^{A}}{{\Delta}^{3/2}}\left[(z-z^{\prime})\cosh Y - t\sinh Y - \mathrm{sgn}(Y) A\sqrt{\Delta} - \frac{\sigma_{0}\sinh Y}{\tanh^2 Y}\Delta \right]  \bigg{\}},\\
        e{E}_{z,\mathrm{P}}^{\pm}(t, \mathbf{r}) &= \alpha_{\mathrm{EM}} \int_{V_{\pm}} \mathrm{d}^3\mathbf{r}^{\,\prime} \int_{-Y_0}^{Y_0} \mathrm{d}Y \ \Psi_{\pm}(Y)\rho_{\pm}(\mathbf{r}^{\,\prime})\tilde{\Theta}_{\mp}(\mathbf{r}_{\perp}^{\,\prime}) \bigg{\{}
        \frac{\mathrm{sgn}(Y)\sigma_{0}^2}{\tanh^2 Y} \exp\left(-\sigma_{0}\left[ t - \frac{z-z^{\prime}}{\tanh Y} \right]\right) \Gamma(0, -A)\\
        \quad\quad\quad&\quad +\frac{e^{A}}{{\Delta}^{3/2}} \left[ (z-z^{\prime})\cosh Y - t\sinh Y - \mathrm{sgn}(Y) A\sqrt{\Delta} - \frac{\sigma_{0}\sinh Y}{\tanh^2 Y}\Delta \right]  \bigg{\}},
        \end{aligned}
        \end{equation}
        where the sign function $\mathrm{sgn}(Y)=Y/|Y|$ denotes the sign of rapidity $Y$. Note that Eqs. (\ref{Max_SP01}-\ref{Max_ST02}) quantify the $x$, $y$ and $z$ components of EM fields contributed by spectators and participants in heavy-ion collisions with a constant electric conductivity $\sigma_{0}$ implemented, where we also incorporate the generalized Lorentz contracted charge distributions, i.e., Eqs. (\ref{3pF_01}) or (\ref{3pF_09}), and the retardation correction through $\Psi_{\pm}(Y)$. On the one hand, we currently in Eqs. (\ref{Max_SP01}-\ref{Max_ST02}) do not include the chiral magnetic conductivity $\sigma_{\chi}$ into the Maxwell's equations in Eq. (\ref{Max_Eq01}) since in general $\sigma_{\chi} \ll \sigma$. Also, the value of electric conductivity $\sigma$ from recent lattice calculations~\cite{Arnol2003,Gupta2004,GAart2007,HDing2011,Franc2013,Aless2013,Brand2013} still shows very large uncertainties, which may easily submerge the contributions from chiral magnetic conductivity $\sigma_{\chi}$. On the other hand, Ref.~\cite{LiHui2016} has shown that even when the chiral magnetic conductivity $\sigma_{\chi}$ is included, the dominative components such as $B_{\phi}$, $E_r$ and $E_z$ will not be changed at all. Therefore, we hold the point of view that Eqs. (\ref{Max_SP01}-\ref{Max_ST02}) can well quantify the dominant contributions of medium feedback effects for the time evolution and centrality (impact parameter $b$) dependence of EM fields generated in the pure conducting medium.

        Here we note that the chiral magnetic conductivity $\sigma_{\chi}$ in general will be complex due to the spatially anti-symmetric part of the off diagonal photon polarization tensor, as has been formally discussed and evaluated in~\cite{Kharzv2009} by using the linear response theory. It has also been shown in~\cite{Kharzv2009} that the chiral magnetic conductivity $\sigma_{\chi}$ has very strong frequency $\omega$ and temperature $T$ dependence, and the induced electric current $j(t)$ can even change from positive to negative during the time evolution. We therefore currently postpone the inclusion of the chiral magnetic conductivity $\sigma_{\chi}$ into the extended KMW model, but leave it to our future study.

\subsection{Electromagnetic Fields from the Extended KMW Model during QGP Evolution}
\label{sec:Electromagnetic Fields from the Extended KMW Model during QGP Evolution}

        It is generally acknowledged in heavy-ion collisions experiments that the QGP can be created at the condition of high collision energies, extremely high temperatures and large energy densities. Such a condition should be satisfied in Au+Au collisions at $\sqrt{s}=200\mathrm{\,GeV}$ at RHIC or in Pb+Pb collisions at $\sqrt{s}=2760\mathrm{\,GeV}$ at LHC. These created deconfined QGP medium can only exist after heavy-ion collisions. Thus, the conducting property of the medium should only exist after the medium time $t_{\sigma}$ ($t_{\sigma}\gtrsim 0$), which is the time that we assume the medium feedback effects start to work.

        In such a sense as we mention above, medium feedback effects should only be considered when the retardation time $t_{\mathrm{ret}}$ is after the medium time $t_{\sigma}$, ahead of which one should consider using the formulas of EM fields in the vacuum. When the conductivity $\sigma$ is switched on after the medium time $t_{\sigma}$, one can roughly approximate the medium effects by using a constant Ohm electric conductivity $\sigma_{0}$ for simplicity. Hence, a time-dependent step-function-like conductivity could be employed, namely $\sigma \to \tilde{\sigma}(t) = \sigma_{0}\cdot \theta(t-t_{\sigma})$. Thus for the generated magnetic field, Eqs. (\ref{Max_SP01}-\ref{Max_PT01}) should be accordingly modified as
        \begin{equation}\label{MaxR_SP01}
        \begin{aligned}
        e\mathbf{B}_{\mathrm{S}}^{\pm}(t, \mathbf{r}) &= \lim_{Y\to \pm Y_0} \alpha_{\mathrm{EM}}\sinh Y \int_{V_{\pm}} \mathrm{d}^3\mathbf{r}^{\,\prime} \frac{ \rho_{\pm}(\mathbf{r}^{\,\prime})\left[1- \tilde{\Theta}_{\mp}(\mathbf{r}_{\perp}^{\,\prime})\right] \mathbf{e}_{z} \times \mathbf{R}_{\pm} }{ \Delta^{3/2} }\left[1 + \frac{\tilde{\sigma}(t_{\mathrm{ret}})\sinh|Y| }{2}\sqrt{\Delta} \right]e^{\tilde{A}},\quad\quad\;\\
        \end{aligned}
        \end{equation}
        \begin{equation}\label{MaxR_PT01}
        \begin{aligned}
        e\mathbf{B}_{\mathrm{P}}^{\pm}(t, \mathbf{r}) &= \alpha_{\mathrm{EM}}\int_{V_{\pm}} \mathrm{d}^3\mathbf{r}^{\,\prime} \int_{-Y_0}^{Y_0} \mathrm{d}Y \frac{ \Psi_{\pm}(Y)\sinh Y \cdot \rho_{\pm}(\mathbf{r}^{\,\prime})\tilde{\Theta}_{\mp}(\mathbf{r}_{\perp}^{\,\prime}) \mathbf{e}_{z} \times \mathbf{R}_{\pm}  }{ \Delta^{3/2} } \left[1 + \frac{\tilde{\sigma}(t_{\mathrm{ret}})\sinh|Y| }{2}\sqrt{\Delta} \right]e^{\tilde{A}},
        \end{aligned}
        \end{equation}
        for the contributions of spectators and participants, with
        \begin{equation}\label{MaxR_SP02}
        \begin{aligned}
        R_{\perp} &\equiv |\mathbf{r}_{\perp}-\mathbf{r}_{\perp}^{\prime}|=\sqrt{ (x-x^{\,\prime})^2 + (y-y^{\,\prime})^2 },\\
        \Delta    &\equiv \left[ t\sinh Y - (z - z^{\prime})\cosh Y \right]^2 + {R_{\perp}}^2,\\
        \tilde{A} &\equiv \frac{\sinh Y }{2}\left[t\sinh Y - (z - z^{\prime})\cosh Y \right] - \frac{\tilde{\sigma}(t_{\mathrm{ret}})\sinh|Y|}{2}\sqrt{\Delta}.\\
        \end{aligned}
        \end{equation}
        Likewise, the generated electric field in Eqs. (\ref{Max_ST01}-\ref{Max_ST02}) should be modified as
        \begin{equation}\label{MaxR_ST01}
        \begin{aligned}
        \begin{pmatrix} e{E}_{x,\mathrm{S}}^{\pm}\\ e{E}_{y,\mathrm{S}}^{\pm} \end{pmatrix}(t, \mathbf{r}) &= \lim_{Y\to \pm Y_0} \alpha_{\mathrm{EM}} \int_{V_{\pm}} \mathrm{d}^3\mathbf{r}^{\,\prime} \rho_{\pm}(\mathbf{r}^{\,\prime})\left[1- \tilde{\Theta}_{\mp}(\mathbf{r}_{\perp}^{\,\prime})\right]\bigg{\{}\frac{ R_{\perp}\cosh Y}{\Delta^{3/2}}\left(1 + \frac{\tilde{\sigma}(t_{\mathrm{ret}})\sinh |Y|}{2}\sqrt{\Delta}\right)\\
        \quad\quad\quad&\quad - \frac{\tilde{\sigma}(t_{\mathrm{ret}})}{R_{\perp} \tanh |Y|} \left[1 + \frac{\sinh |Y|}{\sqrt{\Delta}} \left(t - \frac{z - z^{\prime}}{\tanh Y}\right) \right] \exp\left(-\tilde{\sigma}(t_{\mathrm{ret}})\left[ t - \frac{z-z^{\prime}}{\tanh Y} \right]\right) \bigg{\}}\frac{e^{\tilde{A}}}{R_{\perp}}\begin{pmatrix} x-x^{\prime}\\ y-y^{\prime} \end{pmatrix},\\
        \begin{pmatrix} e{E}_{x,\mathrm{P}}^{\pm}\\ e{E}_{y,\mathrm{P}}^{\pm} \end{pmatrix}(t, \mathbf{r}) &= \alpha_{\mathrm{EM}} \int_{V_{\pm}} \mathrm{d}^3\mathbf{r}^{\,\prime} \int_{-Y_0}^{Y_0} \mathrm{d}Y \ \Psi_{\pm}(Y)\rho_{\pm}(\mathbf{r}^{\,\prime})\tilde{\Theta}_{\mp}(\mathbf{r}_{\perp}^{\,\prime})\bigg{\{}\frac{ R_{\perp}\cosh Y}{\Delta^{3/2}}\left(1 + \frac{\tilde{\sigma}(t_{\mathrm{ret}})\sinh |Y|}{2}\sqrt{\Delta}\right)\\
        \quad\quad\quad&\quad - \frac{\tilde{\sigma}(t_{\mathrm{ret}})}{R_{\perp} \tanh |Y|} \left[1 + \frac{\sinh |Y|}{\sqrt{\Delta}} \left(t-\frac{z - z^{\prime}}{\tanh Y}\right) \right] \exp\left(-\tilde{\sigma}(t_{\mathrm{ret}})\left[ t - \frac{z-z^{\prime}}{\tanh Y} \right]\right)\bigg{\}}\frac{e^{\tilde{A}}}{R_{\perp}}\begin{pmatrix} x-x^{\prime}\\ y-y^{\prime} \end{pmatrix},
        \end{aligned}
        \end{equation}
        for the $x$ and $y$ components of contributions from spectators and participants. For that of $z$ components, we have
        \begin{equation}\label{MaxR_ST02}
        \begin{aligned}
        e{E}_{z,\mathrm{S}}^{\pm}(t, \mathbf{r}) &= \lim_{Y\to \pm Y_0} \alpha_{\mathrm{EM}} \int_{V_{\pm}} \mathrm{d}^3\mathbf{r}^{\,\prime} \rho_{\pm}(\mathbf{r}^{\,\prime})\left[1- \tilde{\Theta}_{\mp}(\mathbf{r}_{\perp}^{\,\prime})\right] \bigg{\{}
        \frac{\mathrm{sgn}(Y) \tilde{\sigma}^2(t_{\mathrm{ret}})}{\tanh^2 Y} \exp\left(-\tilde{\sigma}(t_{\mathrm{ret}})\left[ t - \frac{z-z^{\prime}}{\tanh Y} \right]\right) \Gamma(0, -\tilde{A})\\
        \quad\quad\quad&\quad + \frac{e^{\tilde{A}}}{{\Delta}^{3/2}}\left[(z-z^{\prime})\cosh Y - t\sinh Y - \mathrm{sgn}(Y) \tilde{A}\sqrt{\Delta} - \frac{\tilde{\sigma}(t_{\mathrm{ret}})\sinh Y}{\tanh^2 Y}\Delta \right]  \bigg{\}},\\
        e{E}_{z,\mathrm{P}}^{\pm}(t, \mathbf{r}) &= \alpha_{\mathrm{EM}} \int_{V_{\pm}} \mathrm{d}^3\mathbf{r}^{\,\prime} \int_{-Y_0}^{Y_0} \mathrm{d}Y \ \Psi_{\pm}(Y)\rho_{\pm}(\mathbf{r}^{\,\prime})\tilde{\Theta}_{\mp}(\mathbf{r}_{\perp}^{\,\prime}) \bigg{\{}
        \frac{\mathrm{sgn}(Y)\tilde{\sigma}^2(t_{\mathrm{ret}})}{\tanh^2 Y} \exp\left(-\tilde{\sigma}(t_{\mathrm{ret}})\left[ t - \frac{z-z^{\prime}}{\tanh Y} \right]\right) \Gamma(0, -\tilde{A})\\
        \quad\quad\quad&\quad +\frac{e^{\tilde{A}}}{{\Delta}^{3/2}} \left[ (z-z^{\prime})\cosh Y - t\sinh Y - \mathrm{sgn}(Y) \tilde{A}\sqrt{\Delta} - \frac{\tilde{\sigma}(t_{\mathrm{ret}})\sinh Y}{\tanh^2 Y}\Delta \right]  \bigg{\}}.
        \end{aligned}
        \end{equation}
        Here we note that when the retardation time $t_{\mathrm{ret}}$ is ahead of the medium time $t_{\sigma}$, namely $t_{\mathrm{ret}}< t_{\sigma}$, the conductivity $\tilde{\sigma}$ will vanish and Eqs. (\ref{MaxR_SP01}-\ref{MaxR_ST02}) will naturally reduce to the formulas of EM fields in the vacuum, i.e. Eqs. (\ref{LS_BS05}-\ref{LS_PT03}). Such a consistent property has also been clearly revealed in the point-like charge case, see Eqs. (\ref{Tanfs_Eq02}-\ref{Max_Eq05}). Thus we would like to refer to Eqs. (\ref{MaxR_SP01}-\ref{MaxR_ST02}) as a simplified, ``mixed" estimation of generated EM fields with contributions from both source charges in the vacuum and that in the conducting medium. In the same sense, we can refer to Eqs. (\ref{Max_SP01}-\ref{Max_ST02}) as the estimation of EM fields in the pure conducting medium, and Eqs. (\ref{Tanfs_Eq02}-\ref{Max_Eq05}) as that in the pure vacuum.

        Before we move forward, let us make some short remarks here. The merits of above generalization for the charge number densities as well as the formulations of estimated EM fields both in the pure vacuum in Eqs. (\ref{LS_BS05}-\ref{LS_PT03}), in the pure conducting medium in Eqs. (\ref{Max_SP01}-\ref{Max_ST02}) and during the realistic QGP evolution in Eqs. (\ref{MaxR_SP01}-\ref{MaxR_ST02}) in the extended KMW model can be at least boiled down to two points: one is that it is more realistically akin to the physical situations for charge distributions with relativistic motion, for which both spherical formed and axially deformed charge number densities are explicitly elaborated, e.g., Eqs. (\ref{3pF_01}) or (\ref{3pF_09}). Therefore, the formulations of estimated EM fields based on the generalized charge number densities can be easily applied to various colliding systems, such as the asymmetric $\mathrm{Cu+Au}$ colliding system or the ongoing STAR experiments of isobar colliding systems in which the $\mathrm{^{96}Zr}$ and $\mathrm{^{96}Ru}$ nuclei are widely regarded as axially deformed nuclei~\cite{WDeng2017,VKoch2017,Huang2017,SZShi2018}. Here we note that a lot of discussions and the related estimations for the EM fields generated in isobar collisions have been presented~\cite{XLZhaC2019,Hamelm2019,WDeng2017,Huang2017,haoXu2018,Jiang2018,SZShi2019}.

        Meanwhile, since the charge number density $\rho(\mathbf{r}^{\prime})$ in Eqs. (\ref{3pF_01}) or (\ref{3pF_09}) is formally embedded with relativistic effect due to Lorentz contraction, it will be much more appropriate and adaptable for the situation at lower energy regions where the Lorentz contraction effect is not that significantly large, such as at lower RHIC energy with $\sqrt{s}=7.7\mathrm{\,GeV}$ where the Lorentz factor $\gamma\sim 4$. As a comparison, the ``pancake shape disk" approximation used in~\cite{Kharzv2008} will only be a good approximation in high energy regions where the Lorentz factor $\gamma$ is substantially large. Therefore the extended KMW model proposed in this paper will be much more appropriate for applications to the STAR-BES program, and some even lower energy regions like the under planning FAIR, NICA and J-PARC programs.

        The other point is that since the estimation of generated EM fields are explicitly embedded with medium feedback effects through a constant electric conductivity $\sigma_{0}$ in the pure conducting medium and a time-dependent conductivity $\tilde{\sigma}(t)$ during the realistic QGP evolution along with the refined baryon-junction stopping effect with proper retardation correction in Eq. (\ref{LS_PT03}), the generalized formulas of EM fields from Eqs. (\ref{LS_BS05}-\ref{LS_PT03}) in the pure vacuum, or Eqs. (\ref{Max_SP01}-\ref{Max_ST02}) in the pure conducting medium, or Eqs. (\ref{MaxR_SP01}-\ref{MaxR_ST02}) during the realistic QGP evolution hence can further be used for many EM fields related studies, such as the CME related charge asymmetry fluctuations or correlators~\cite{Kharzv2008,Selyu2006,SZShi2018,Jiang2018}, the CME vector current $\mathbf{J}_{\chi}=\sigma_{\chi}\mathbf{B}$ and its related studies for chiral magnetic conductivity $\sigma_{\chi}$~\cite{Kharzv2009}, modifications of in-medium particle's mass~\cite{HaoLei2016,Bonati2017,GSBali2018,Coppol2018,Shijun2019,Ferrer2019,HTDing2020} as well as the QCD phase diagram under strong magnetic field~\cite{Stefa2018,Masim2018,GShao2019,Shile2020}, and so on.

\section{Results and Discussions}
\label{sec:Results and Discussions}

        Since we have explicitly formulated the estimations of generated EM fields for heavy-ion collisions in the vacuum in Eqs. (\ref{LS_BS05}-\ref{LS_PT03}), in the pure QGP medium in Eqs. (\ref{Max_SP01}-\ref{Max_ST02}) and also during the realistic QGP evolution in Eqs. (\ref{MaxR_SP01}-\ref{MaxR_ST02}), as well as an alternative solution to doing Monte-Carlo simulations during the realistic QGP evolution in Eqs. (\ref{Appex_LW01}-\ref{Appex_LW03}), let us first give some pre-analysis before performing the $ab$ $initio$ integration of charge distribution or event-by-event simulations of generated EM fields at specific space-time points. Due to the mirror and centrosymmetric symmetries of the colliding system (as is illustrated in Fig. \ref{Fig01_CS}), the total electric field $\mathbf{E}$ will vanish at the center point $\mathbf{r}=\mathbf{0}$ of the overlapping region while the total magnetic field $\bf{B}$ will only remain its $y$ component $B_y$ pointing in the negative $y$ direction, which may further help to result in the charge asymmetry with respect to reaction plane due to the CME. Therefore, we mainly focus on the $y$ component of magnetic field $B_y$ in this paper.

        Let us now declare some abbreviations that we will present in the following. The original KMW model is incorporated with a two-dimensional (2D) surface charge number density, which is abbreviated as ``2D KMW". The extended KMW model is generalized with a three-dimensional (3D) volume charge number density, which is likewise abbreviated as ``3D KMW" including two cases: with (w/ ) and without (w/o) retardation correction (RC). The pure vacuum case and pure medium case is abbreviated as ``Pure Vac." and ``Pure Med.", such that the case during the realistic QGP evolution in Eqs. (\ref{MaxR_SP01}-\ref{MaxR_ST02}) is abbreviated as ``Mixed".  The collision time $t_{c}$ and medium time $t_{\sigma}$ for simplicity are chosen at fully overlapping time, namely $t_c=t_{\sigma}=0$.

        Besides, we do check that even when one considers the collision time $t_{c}$ in Eq. (\ref{LS_PT03}) totally from the geometry consideration regardless of the causality reason according to the theory of special relativity, namely the collision time $t_c$ can be estimated as $t_c=-t_{d}$, where $t_{d}$ is the departure time defined in following Eq. (\ref{Res_01}), our results basically will not be changed due to the much earlier retardation time $t_{\mathrm{ret}}$.

        It should also be noted that the formulation of estimated magnetic field in the original KMW model is actually only limited in the vacuum scenario without considering medium effects. In order to make a better comparison of the extended KMW model with the original KMW model in the pure conducting medium, one can also obtain the generalized 2D KMW model embedded with a constant electric conductivity $\sigma_{0}$ from the reduction of extended KMW model by removing the longitudinal position dependence and retardation correction, and replacing the corresponding charge number density. Such a generalization of magnetic field in the pure conducting medium is actually consistent with that in Refs.~\cite{Gursoy2014,Gursoy2018}.

        Moreover, it is mainly the magnetic field strength that has been numerically evaluated in the original KMW model~\cite{Kharzv2008} and also in the following papers~\cite{Gursoy2014,Gursoy2018}, thus we will only show the results of estimated magnetic field in this paper for the sake of comparison. The results of estimated electric field can be similarly evaluated from the extended KMW model in the same way as we present here.

\subsection{Time Evolution of Magnetic Field Strength}
\label{sec:Time Evolution of Magnetic Field Strength}

\begin{figure}[htb]
        \includegraphics[angle=0,scale=0.48]{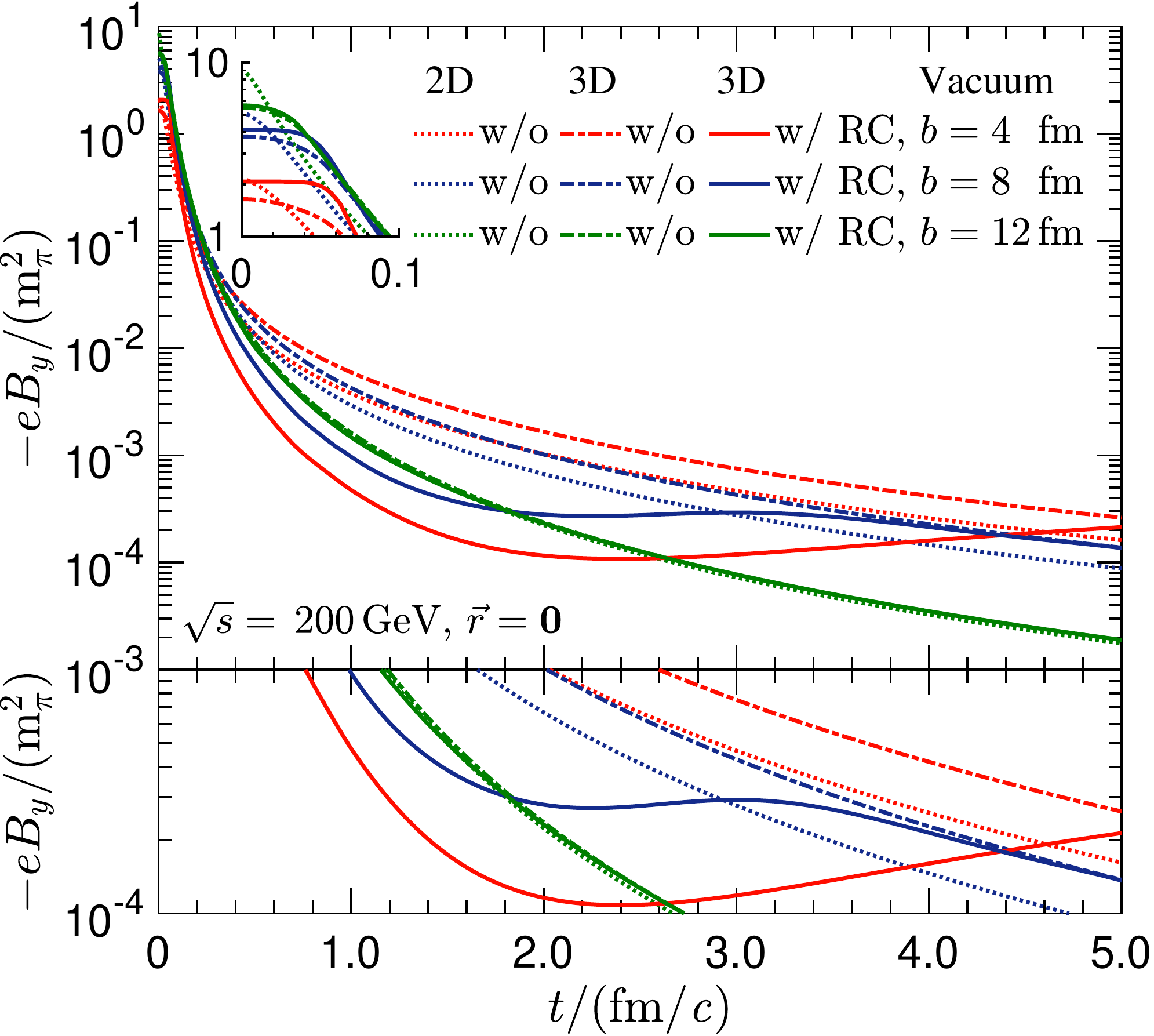}
        \caption{(Color online) Comparison of the time evolution of the total magnetic field strength in the pure vacuum from the extended KMW model with that from the original KMW model at different impact parameters $b=4,\,8,\,12$ fm with (w/ ) and without (w/o) retardation correction (RC) for the center point $\mathbf{r}=\mathbf{0}$ of the overlapping region in Au+Au collisions at $\sqrt{s}=200\mathrm{\,GeV}$.}
        \label{Fig02_eBTime}
        \end{figure}

        Let us now make a sample comparison of the time evolution of the total magnetic field strength estimated from Eq. (\ref{LS_BS05}) in the vacuum. In Fig. \ref{Fig02_eBTime}, we show the estimation of time evolution of the total magnetic field strength $e\mathbf{B}(t,\mathbf{r})$ in the vacuum at the center point $\mathbf{r}=\mathbf{0}$ of the overlapping region in Au+Au collisions at $\sqrt{s}=200\mathrm{\,GeV}$ with different impact parameters, i.e. $b=4,\,8,\,12$ fm. From the inserted plot in the upper panel, one can clearly see that the magnetic field from the 3D KMW model with and without retardation correction at $t=0$ will systematically yield a relatively smaller magnetic field strength especially for larger impact parameter $b$, compared with that too sharp cusps of magnetic field strength around $t\sim 0$ from the original KMW model. It can intuitively attributed to the more localized and central elevated charge distribution used in the original KMW model. We check that our results at $t=0$ are actually consistent with the numerical simulations from HIJING model in~\cite{DenHu2012}, which will be clearly redemonstrated in the next subsection.

        However, compared with that ``seemly enhanced" magnetic field strength from the extended KMW model without retardation correction, it is also noticeable that the original KMW model will generally give rise to a weaker magnetic field strength just shortly ($t\sim 0.03\mathrm{\,fm/{\it c}}$) after the collision. Moreover, the differences in between become rather noticeable at later-time stages for more central (or smaller impact parameter $b$) collisions. Thus the extended KMW model without retardation correction will yield a longer lifetime $t_{\mathrm{B}}$ of the magnetic field than the original KMW model. When incorporated with retardation correction, however, the magnetic field strength from the extended KMW model will firstly decrease very fast, and then decrease slowly and even grow up to a certain extent for some time range at some smaller impact parameters, e.g., $b=4,\,8\mathrm{\,fm}$, as clearly shown in the lower panel of Fig. \ref{Fig02_eBTime}. This can be intuitively understood as follows: when incorporated with the retardation correction in Eq. (\ref{LS_PT03}), the baryon-junction stopping effects represented through $f_{\pm}(Y)$ in Eq. (\ref{LS_PT03}) are basically ruled out by the $\theta(t_{\mathrm{ret}}-t_c)$ function during the comparison of retardation time $t_{\mathrm{ret}}$ with the collision time $t_c$. The contributions of charged participants are mainly dominated by those with beam rapidity $\pm Y_{0}$, hence the magnetic field strength at early-time stage decreases much faster than those without retardation correction; At middle- or later-time stages, however, the contributions of baryon-junction stopping bulk matter grows dominant since the spectators have flown away. The resulting magnetic field strength contributed by those participants will increase first and then decrease. Hence the resulting total magnetic field strength will show non-monotonic behaviors, especially for smaller impact parameter $b$ due to larger baryon-junction stopping bulk matter. This can be interpreted as a result of the competition between retardation effect of field propagation and baryon-junction stopping effect of participants.

\begin{figure}[htb]
        \includegraphics[angle=0,scale=0.48]{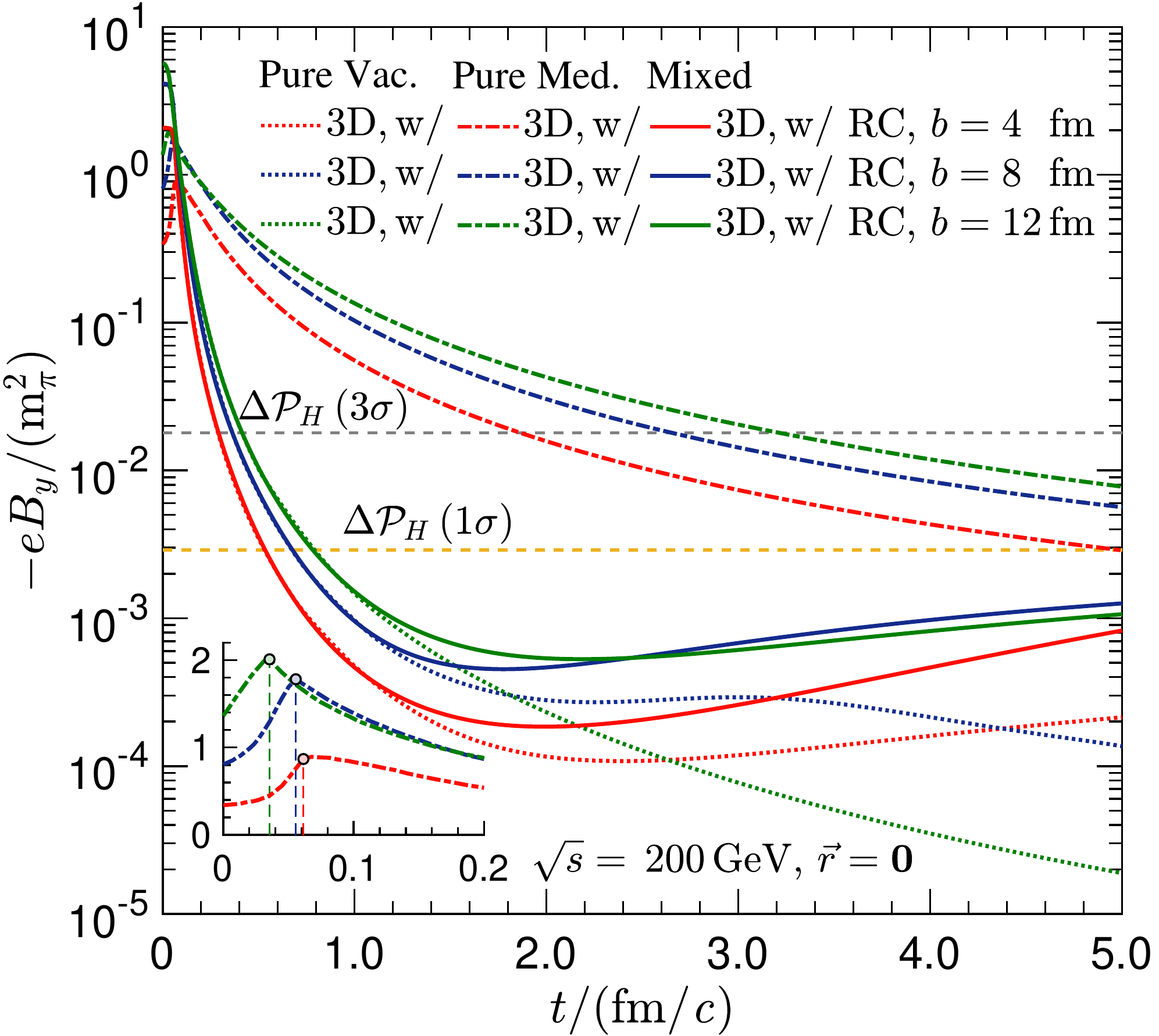}
        \caption{(Color online) Comparison of the time evolution of the total magnetic field strength during the realistic QGP evolution with a time-dependent electric conductivity $\tilde{\sigma}(t)=\sigma_{0}\cdot\theta(t-t_{\sigma})$ (denoted as ``Mixed") from the extended KMW model with that in the pure vacuum in Fig. \ref{Fig02_eBTime} (denoted as ``Pure Vac.") and that in the pure conducting medium (denoted as ``Pure Med.") at different impact parameters $b=4,\,8,\,12$ fm for the center point $\mathbf{r}=\mathbf{0}$ in Au+Au collisions at $\sqrt{s}=200\mathrm{\,GeV}$. Two horizontal dashed lines indicate the later-time constraints of the magnetic field strength estimated in~\cite{Berndt2018} from the difference between global polarization difference of $\Lambda$ and $\bar{\Lambda}$ hyperons in Au+Au collisions at $\sqrt{s}=200\mathrm{\,GeV}$ in~\cite{JAdam2018} at one standard deviation ($1\sigma$) and three standard deviations ($3\sigma$), respectively.}
        \label{Fig03_eBTimeMixed}
        \end{figure}

        The electric conductivity $\sigma_{0}$ has been calculated by using a first principal lattice QCD approach in the quenched approximation in Ref.~\cite{HDing2011}, according to which we set the electric conductivity $\sigma_{0}=5.8\mathrm{\,MeV}$ in this paper so as to make estimations of the magnetic field with medium feedback effects. In Fig. \ref{Fig03_eBTimeMixed}, we show the time evolution of total magnetic field strength during the realistic QGP evolution with a time-dependent electric conductivity $\tilde{\sigma}(t)=\sigma_{0}\cdot\theta(t-t_{\sigma})$ (denoted as ``Mixed") from the extended KMW model in Eqs. (\ref{MaxR_SP01}-\ref{MaxR_ST02}) at the center point $\mathbf{r}=\mathbf{0}$ of the overlapping region in Au+Au collisions at $\sqrt{s}=200\mathrm{\,GeV}$ with different impact parameters $b=4,\,8,\,12$ fm. The mixed results have contributions both coming from charged sources in the vacuum and that in the pure conducting medium. To make a better comparison, we also show the corresponding results in the pure vacuum in Fig. \ref{Fig02_eBTime} (denoted as ``Pure Vac.") and that in the pure conducting medium (denoted as ``Pure Med.").

        In Fig. \ref{Fig03_eBTimeMixed}, one can clearly see that the magnetic field strength in the pure conducting medium grows up at early-time stages while decreases slowly at later-time stages. Thus, there clearly show non-monotonic behaviors at early-times stages around $t\sim 0.05 \mathrm{\,fm/{\it c}}$ in the pure conducting medium. However, compared with these pure medium results, the mixed results for a realistic QGP evolution show drastically different behaviors, namely the monotonically decreasing behaviors at early-time stages while slowly enhanced behaviors at later-time stages. Meanwhile, we notice that the differences between the solid lines (mixed results) and the corresponding dotted lines (pure vacuum results) at early-time stages are almost indistinguishable. Therefore, we realize that the magnetic field strength during the realistic QGP evolution at early-time stages are mostly contributed by charged sources in the pure vacuum, and the medium feedback effects are almost negligible at this time range ($t\leq 1.2 \mathrm{\,fm/{\it c}}$) during the realistic QGP evolution, for which the strength of magnetic field has decreased down to below $\sim 10^{-2} m_{\pi}^2$. Hence, one may conclude that the pure medium case is not an appropriate approximation for the generation of magnetic field at early-time stages in heavy-ion collisions, which will overestimate the magnetic field strength by orders of magnitude.

        Moreover, we notice that the time evolution of the estimated magnetic field strength in the pure conducting medium at later-time stages in Fig. \ref{Fig03_eBTimeMixed} is quiet different from that in the vacuum case for the impact parameter $b$ dependence. This can be intuitively understood as follows: since in the vacuum case there is no medium feedback effect due to the vanishing electric conductivity, the magnetic field decays very fast and there is only the baryon-junction stopping effect that may to some extent hinder the moving charged bulk matter from flying away from the field point $\mathbf{r}=\mathbf{0}$. Thus the baryon-junction stopping effect will dominate in the later-time stages especially for smaller impact parameter $b$ due to larger charged bulk matter of participants, the resulting total magnetic field strength therefore becomes larger for smaller impact parameter $b$ at later-time stages. In the pure conducting medium, however, since the electric conductivity $\sigma_{0}$ is always constant and non-negligible, it is the Faraday's induction effect rather than the baryon-junction stopping effect that dominates for the later-time stages evolution. The resulting total magnetic field strength in the pure conducting medium therefore shows rather distinct impact parameter $b$ dependence at later-time stages, namely larger impact parameter $b$ corresponds to stronger total magnetic field strength ($b\leq 12\mathrm{\, fm}$).

        Compared with pure vacuum results and pure medium results, the mixed results for a realistic QGP evolution lie in between them, receiving together effects from both baryon-junction stopping effect and medium feedback effects. Thus, the enhancements of magnetic field strength at later-time stages become rather significant than those pure vacuum results due to more dominative medium feedback effects, which can be clearly reflected in the differences between pure vacuum results and the corresponding mixed results, further indicating that the resulting magnetic field strength receives growing larger contributions from charged sources in the conducting QGP medium and the pure vacuum scenario is not appropriate at later-time stages QGP evolution.

        More specifically during the QGP evolution, the mixed results at $ 0.1<t<2.0 \mathrm{\,fm/{\it c}}$ are always much smaller than the pure medium results, indicating that pure medium scenario is not realistic and appropriate enough for early-time stages QGP evolution in heavy-ion collisions. For even later-time stages, e.g. $ 2.0<t<5.0 \mathrm{\,fm/{\it c}}$, one can notice that the mixed results are asymptotically growing to the results in the pure conducting medium, indicating that the pure medium scenario may be an appropriate approximation for generated magnetic field at even later-time stages. At $3.0<t<5.0 \mathrm{\,fm/{\it c}}$, we indeed check that for smaller impact parameter $b$ the mixed results are mainly contributed by participants in the conducting medium that dominantly enhance the magnetic field strength. For larger impact parameter $b$, however, the mixed results are mainly contributed by spectators in the conducting medium. Also, we notice that the contributions of spectators in the pure conducting medium start to appear earlier for larger impact parameter $b$, where the starting time can be roughly estimated as $t \simeq R_{\mathrm{A}}-b/2$. Some of these features of the magnetic field strength in Fig. \ref{Fig03_eBTimeMixed} are similar to the results in Refs.~\cite{TuchiR2013,McLerr2014,Tuchin2015,Stewa2018}.

        The ``kinks" marked as the points in the inserted plot of Fig. \ref{Fig03_eBTimeMixed} at different impact parameters $b$ for the pure conducting medium scenario are due to the fact that the two colliding nuclei start to separate from each other. The corresponding vertical dashed lines indicate the departure time $t_d$ when the two colliding nuclei just depart from each other, which can be estimated from the following geometry relation in the $x-z$ plane (RP) at a given impact parameter $b$ as follows
        \begin{equation}\label{Res_01}
        \begin{aligned}
        t_d=t_d(b) \equiv   \frac{R_{\mathrm{A}}}{\gamma}\sqrt{1-\left(\frac{b}{2R_{\mathrm{A}}}\right)^2 }.
        \end{aligned}
        \end{equation}
        Thus, there will no longer show any kink at $t\gtrsim R_{\mathrm{A}}/\gamma$ during the time evolution. However, there still exist possible kinks at $t=0$ and $b=2R_{\mathrm{A}}$, since the two colliding nuclei will just missed each other at overlapping time when the observation time $t$ coincides with the departure time $t_d$, namely $t=t_d(b=2R_{\mathrm{A}})=0$.

        By using the interesting idea proposed in the recent Ref.~\cite{Berndt2018}, one can use the experimentally measured polarization data~\cite{Adamc2017,JAdam2018} on $\Lambda$ and $\bar{\Lambda}$ hyperons to make some rough constraints on the allowed later-time magnetic field strength, which has also been recently explored in Ref.~\cite{YuGuo2019}.  It has been estimated in~\cite{Berndt2018} that in the one standard deviation ($1\sigma$) limit of recent measurements on global polarizations of $\Lambda$ and $\bar{\Lambda}$ hyperons in Au+Au collisions at $\sqrt{s}=200\mathrm{\,GeV}$~\cite{JAdam2018}, the magnetic field strength at thermal freeze-out time should be smaller than the $1\sigma$ bound, which follows as
        \begin{equation}\label{Res_02}
        \begin{aligned}
        e|B| &= \frac{eT_s|\Delta \mathcal{P}_{H}|}{2|\mu_{\Lambda}|} < 2.8985 \times 10^{-3}\,m_{\pi}^2,
        \end{aligned}
        \end{equation}
        where $T_s\approx 150\mathrm{\,MeV}$ is the temperature of the emitting source, $\Delta \mathcal{P}_{H} = \mathcal{P}_{\Lambda}-\mathcal{P}_{\bar{\Lambda}}$ is the difference in global polarizations of $\Lambda$ and $\bar{\Lambda}$ hyperons, and $\mu_{\Lambda}=-\mu_{\bar{\Lambda}}=-0.613\mu_{N}$. In the three standard deviations ($3\sigma$) limit, the magnetic field strength should follow the condition: $e|B|<1.8021\times 10^{-2}\,m_{\pi}^2$~\cite{Berndt2018}. Here we choose $m_{\pi}$ as the mass of $\pi^0$ meson rather than that of $\pi^{\pm}$, which is consistent with the pion mass that we use throughout this paper. Therefore, one can immediately use these possible constraints to compare with the magnetic field strength estimated from the extended KMW model during the realistic QGP evolution in Fig. \ref{Fig03_eBTimeMixed}.

        By comparing with possible constraints shown by two horizontal dashed lines at $1\sigma$ and $3\sigma$ bounds in Fig. \ref{Fig03_eBTimeMixed}, we notice that the magnetic field strength in the pure vacuum scenario meets and appears extremely smaller than the $1\sigma$ bound by almost ($1-2$) orders of magnitude at a possible lifetime $t_s \simeq 5{\mathrm{\,fm}/c}$~\cite{HChao2008} of the QGP from hydrodynamic simulations in Au+Au collisions at $\sqrt{s}=200\mathrm{\,GeV}$. Meanwhile, we find that the magnetic field strength in the pure conducting medium with a constant electric conductivity $\sigma_{0}=5.8\mathrm{\,MeV}$ at $3 < t < 5{\mathrm{\,fm}/c}$ almost lies between the $1\sigma$ and $3\sigma$ bounds. At thermal freeze-out time $t=t_s \simeq 5{\mathrm{\,fm}/c}$, we find that the magnetic field strength in the pure conducting medium at impact parameter $b=4-12\mathrm{\,fm}$ lies between $eB\simeq (2.7094-7.8672)\times10^{-3}\,m_{\pi}^2$, which meets the $3\sigma$ bound while violates the $1\sigma$ bound, potentially indicating that the pure conducting medium scenario seems to overestimate the magnetic field strength. For a relatively realistic QGP evolution, the mixed results of the magnetic field strength can meet and approximately approach the $1\sigma$ bound constrained by the global polarizations of $\Lambda$ and $\bar{\Lambda}$ hyperons at $\sqrt{s}=200\mathrm{\,GeV}$.  It may indicate that the mixed case approximation of realistic QGP evolution is more appropriate than the pure conducting medium approximation.
        
        For above comparison with experimental constraints from polarization difference of $\Lambda$ and $\bar{\Lambda}$ hyperons at $\sqrt{s}=200\mathrm{\,GeV}$, we have to clarify that the magnetic field may not be the only possible explanation. Thus, we expect at best our model predictions might describe the polarization difference semi-quantitatively at top RHIC energy. Besides, we have neglected the temperature dependence of the electric conductivity $\sigma$. In lattice QCD simulations, the electric conductivity $\sigma$ should have strong medium temperature $T$ dependence, which has been parameterized and explored in Ref.~\cite{Zahaov2014}. Since it is known that the temperature of QGP medium is decreasing when the QGP fireball is expanding. Therefore, the electric conductivity $\sigma$ should also evolve with time during the realistic QGP evolution. For such a complicated case of magnetic field with thermal background, we can only leave it to our future study.

\subsection{Impact Parameter Dependence of Magnetic Field Strength}
\label{sec: Impact Parameter Dependence of Magnetic Field Strength}

        Let us now compare the impact parameter $b$ dependence of magnetic field strength in the pure vacuum and also during the realistic QGP evolution.

        \begin{figure}[htb]
        \includegraphics[angle=0,scale=0.48]{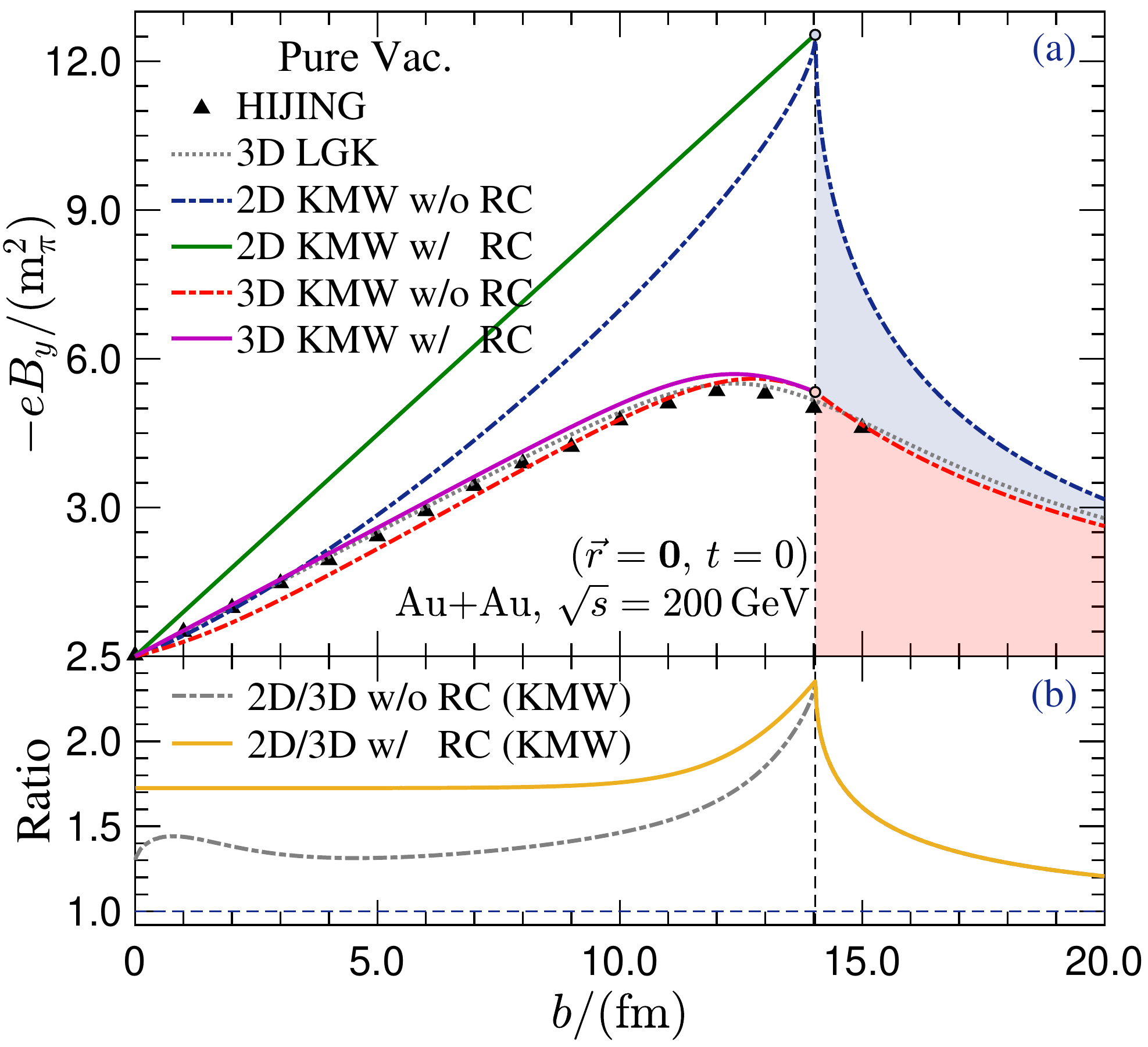}
        \caption{(Color online) (a) Comparison of impact parameter $b$ dependence of the total magnetic field strength in the vacuum at the center point $\mathbf{r}=\mathbf{0}$ of the overlapping region in Au+Au collisions at $\sqrt{s}=200\mathrm{\,GeV}$ at $t=0$ fm/$c$ from six different approaches. (b) The corresponding ratio (2D/3D) of the total magnetic field strength from 2D and 3D KMW models with (w/ ) and without (w/o) retardation correction (RC). The vertical dashed line indicates the boundary at $b=2R_{\mathrm{A}}$, above which the two colliding nuclei will miss each other and there is no contribution from participants, as indicated by the colored bands. The corresponding ``kinks" are also marked by the points along the vertical dashed line.}
        \label{Fig04_eBCen}
        \end{figure}

        In Fig. \ref{Fig04_eBCen}, we show the impact parameter $b$ dependence of the estimated magnetic field strength in the vacuum from the extended KMW model and that from the original KMW model with and without retardation correction at the center point $\mathbf{r}=\mathbf{0}$ of the overlapping region in Au+Au collisions at $\sqrt{s}=200\mathrm{\,GeV}$ at $t=0$. Meanwhile, we also show the results from numerical simulations with HIJING model in~\cite{DenHu2012,Koichi2017} and that from a sample analytical approach (denoted as LGK model) initially proposed in~\cite{Ypeng2014}, both of which have been used for the comparison of impact parameter dependence of magnetic field in Ref.~\cite{Koichi2017}.

        At first glance, one may notice that the magnetic field strength in the vacuum case at smaller impact parameter $b$ from the 3D KMW model seems to deviate a lot from the results obtained with HIJING model in~\cite{DenHu2012}. Meanwhile, we notice that the LGK model seems to work well and be consistent with the HIJING results. Here we should point out however that when using the HIJING model in~\cite{DenHu2012} to calculate the generated magnetic field strength at $t=0$ when the two colliding nuclei are completely overlapping with each other, all nucleons inside the two colliding nuclei are assigned with the same beam velocity $v_z^2 = 1- (2m_{\mathrm{N}}/\sqrt{s})^2$, and a similar treatment has also been employed in the LGK model~\cite{Koichi2017} that all moving charges inside the two colliding nuclei are regarded as spectators. Both two methods neglect the baryon-junction stopping effect during the overlapping process. Unlike these two approaches, we have included the baryon-junction stopping effect through the experimentally supported $f_{\pm}(Y)$ in Eq. (\ref{LS_PT01}) in the extended KMW model, along with the generalized charge distribution. Therefore, it is not surprising that the simulations with HIJING model and the analytic LGK model in~\cite{Koichi2017} can roughly yield very similar results, and both two approaches yield slightly larger magnetic field strength especially at smaller impact parameter $b$ due to larger charged bulk matter of participants affected by $f_{\pm}(Y)$, compared with that from the extended KMW model without retardation correction. When the retardation correction is included, however, the extended KMW model with retardation correction give a slightly larger magnetic field strength at $b\leq 14\mathrm{\,fm}$ than that from the HIJING and LGK approaches, which may be mainly attributed to the generalized charge distribution for the incorporation of Lorentz contraction effect on the geometries of the two colliding nuclei used in the extended KMW model.

        In Fig. \ref{Fig04_eBCen}, one may notice the ``kinks" at $b=2R_{\mathrm{A}}$ from both original and extended KMW models. That is because the two colliding nuclei begins to miss each other at $t=0$ and there is no contribution from participants (which will be demonstrated in the following Fig. \ref{Fig05_eBCenSP}), compared with the other two models that seem to smoothly across the boundary. The ``kinks" are due to the fact that the charge number densities in both 2D and 3D KMW models are accompanied with the step functions $\Theta_{\pm}$ so as to constrain that the charges are limited within the two colliding nuclei (with radius $R_{\mathrm{A}}$ before the Lorentz contraction), e.g. Eqs. (\ref{3pF_01}) or (\ref{3pF_09}). Here we note that the HIJING model uses a relatively large upper nucleus radius limit for the corresponding sampled protons inside, which actually largely exceeds the generally supposed nucleus charge radius $R_{c}$ estimated from the nucleus root mean square (RMS) charge radius $R_{\mathrm{RMS}}$ measured in experiments~\cite{Angel2013}, i.e. $R_{\mathrm{A}} \simeq R_{c} = \sqrt{{5}/{3}}R_{\mathrm{RMS}}$. Similarly, the LGK model used in~\cite{Koichi2017} also employs a relatively large nuclear charge radius $R_{c}$ in the treatment of effective charge $Z_{\mathrm{eff}}$.
        
        \begin{figure}[htb]
        \includegraphics[angle=0,scale=0.48]{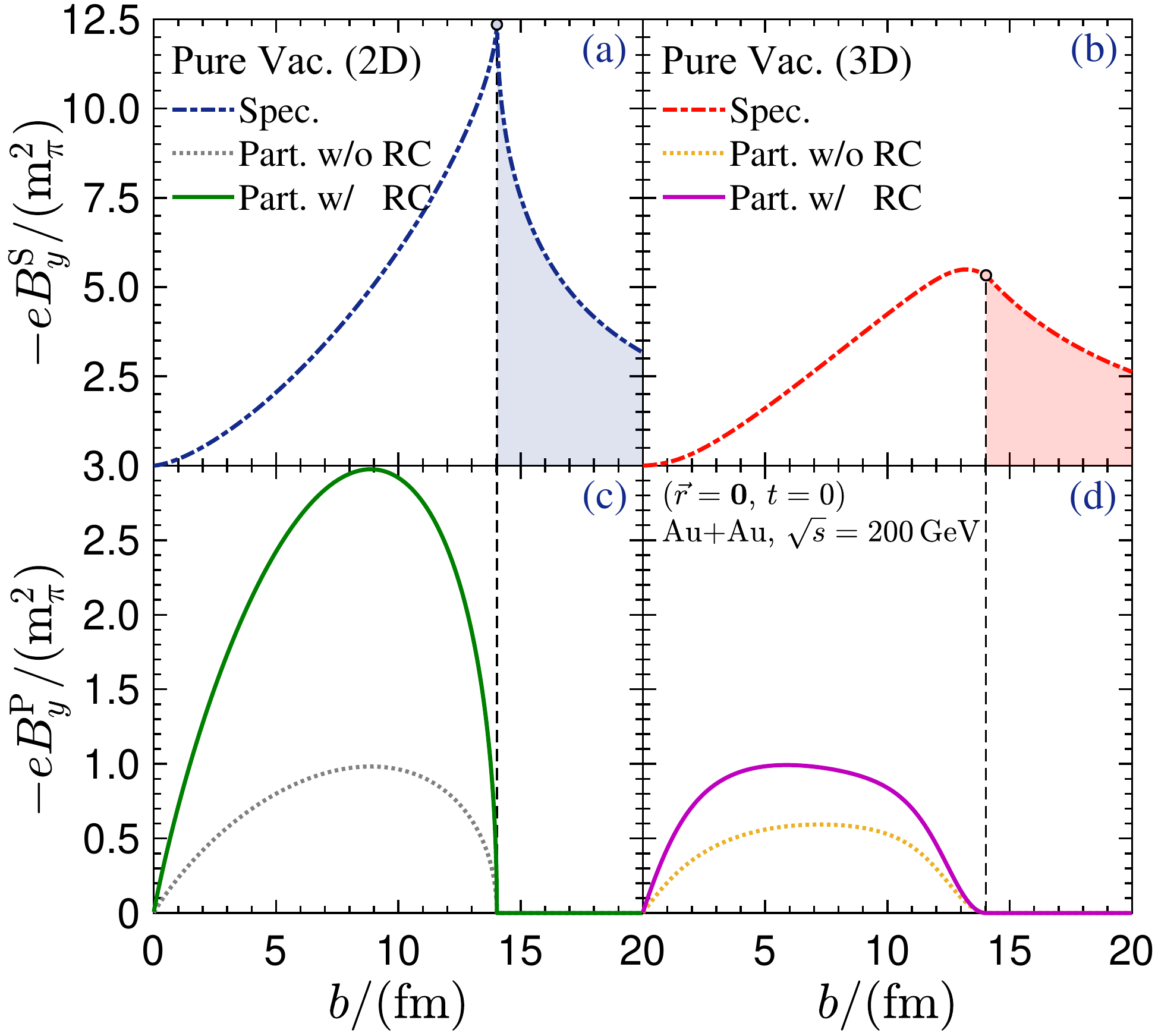}
        \caption{(Color online) Comparison of impact parameter $b$ dependence of the magnetic field strength in the pure vacuum contributed by spectators (S) in the upper panels (a) and (b), and that by participants (P) in the lower panels (c) and (d) at the center point $\mathbf{r}=\mathbf{0}$ of the overlapping region in Au+Au collisions at $\sqrt{s}=200\mathrm{\,GeV}$ at $t=0$ $\mathrm{fm/{\it c}}$ from original (2D) KMW model (left panels) with that from extended (3D) KMW model (right panels). The vertical dashed lines indicate the boundary at $b=2R_{\mathrm{A}}$, above which the two colliding nuclei will miss each other and there is no contribution from participants, as indicated by the colored bands. The corresponding ``kinks" are also marked by the points along the vertical dashed lines.}
        \label{Fig05_eBCenSP}
        \end{figure}

        Besides, one can clearly notice that, compared with other three 3D models used in Fig. \ref{Fig04_eBCen}, the original KMW model indeed largely overestimates the magnetic field strength at $t=0$, which corresponds to the too sharp cusps of magnetic field strength around $t\sim 0$ in Fig. \ref{Fig02_eBTime} especially at larger impact parameter $b$. This can be roughly attributed to the more localized and central elevated 2D surface charge number density and no longitudinal position dependence of the EM fields in the original KMW model. Moreover, we show the ratios (2D/3D) of the total estimated magnetic field strength from the original KMW model to that from the extended KMW model with and without retardation correction in the lower panel of Fig. \ref{Fig04_eBCen}. One can clearly see that the original KMW overestimates the magnetic field strength roughly by a factor of $\sim1.7$ ($1.5$) at relatively smaller impact parameter $b$ with (without) retardation correction, and by a factor of $\sim 2.4$ ($2.4$) at the boundary $b=2R_{\mathrm{A}}$ in particular.

        In Fig. \ref{Fig05_eBCenSP}, we make detailed comparisons of the impact parameter $b$ dependence of the estimated magnetic field strength in the pure vacuum contributed by spectators and participants from the original KMW model with that from the extended KMW model both with and without retardation correction. From the upper panels of Fig. \ref{Fig05_eBCenSP}, one can clearly see that it is mainly the contributions of spectators in the original KMW model that cause the too shape peak of the estimated magnetic field strength, compared with that from the extended KMW model.

        In the lower panels of Fig. \ref{Fig05_eBCenSP}, we also notice that the magnetic field strength contributed by participants with retardation correction systematically shows significant enhancements relative to that without retardation correction for both 2D and 3D KMW models. This is mainly due to the fact that the baryon-junction stopping effect represented by $f_{\pm}(Y)$ in Eq. (\ref{LS_PT03}) is basically ruled out at $t=0$ by the $\theta(t_{\mathrm{ret}}-t_c)$ function during the comparison between retardation time $t_{\mathrm{ret}}$ and collision time $t_c$. The contributions of charged participants are actually all calculated with beam rapidity $\pm Y_{0}$ instead of the normalized rapidity distributions $f_{\pm}(Y)$. Hence this kind of enhancement is quite significant at $t=0$ in Fig. \ref{Fig05_eBCenSP} (c) and (d), especially for the 2D KMW model due to the more localized 2D surface charge number density. It is quite noticeable that the magnetic field strength contributed by participants in the 3D KMW model for a middle impact parameter $b$, e.g. $b=8$ fm, is almost one third of that in the 2D KMW model for the case with retardation correction.

        Different from the situation in the pure vacuum in Fig. \ref{Fig04_eBCen}, the centrality dependence of the generated magnetic field during the QGP evolution in heavy-ion collisions should be realistically evaluated at some time after the collision such that medium feedback effects can indeed take effect on the generated magnetic field. In view of Fig. \ref{Fig03_eBTimeMixed}, we notice that the mixed results and the pure vacuum results are almost indistinguishable at $t\leq1.2\mathrm{\,fm}/c$, thus it may be better to show the centrality dependence at later times during the QGP evolution so as to demonstrate the QGP medium response. Hence we choose to show our results at $t=3\tau_{0}$, where $\tau_{0}=0.6\mathrm{\,fm}/c$ is the widely used (pre-) equilibration time in hydrodynamical simulations~\cite{HChao2008,Peter2003}.
        
        \begin{figure}[htb]
        \includegraphics[angle=0,scale=0.48]{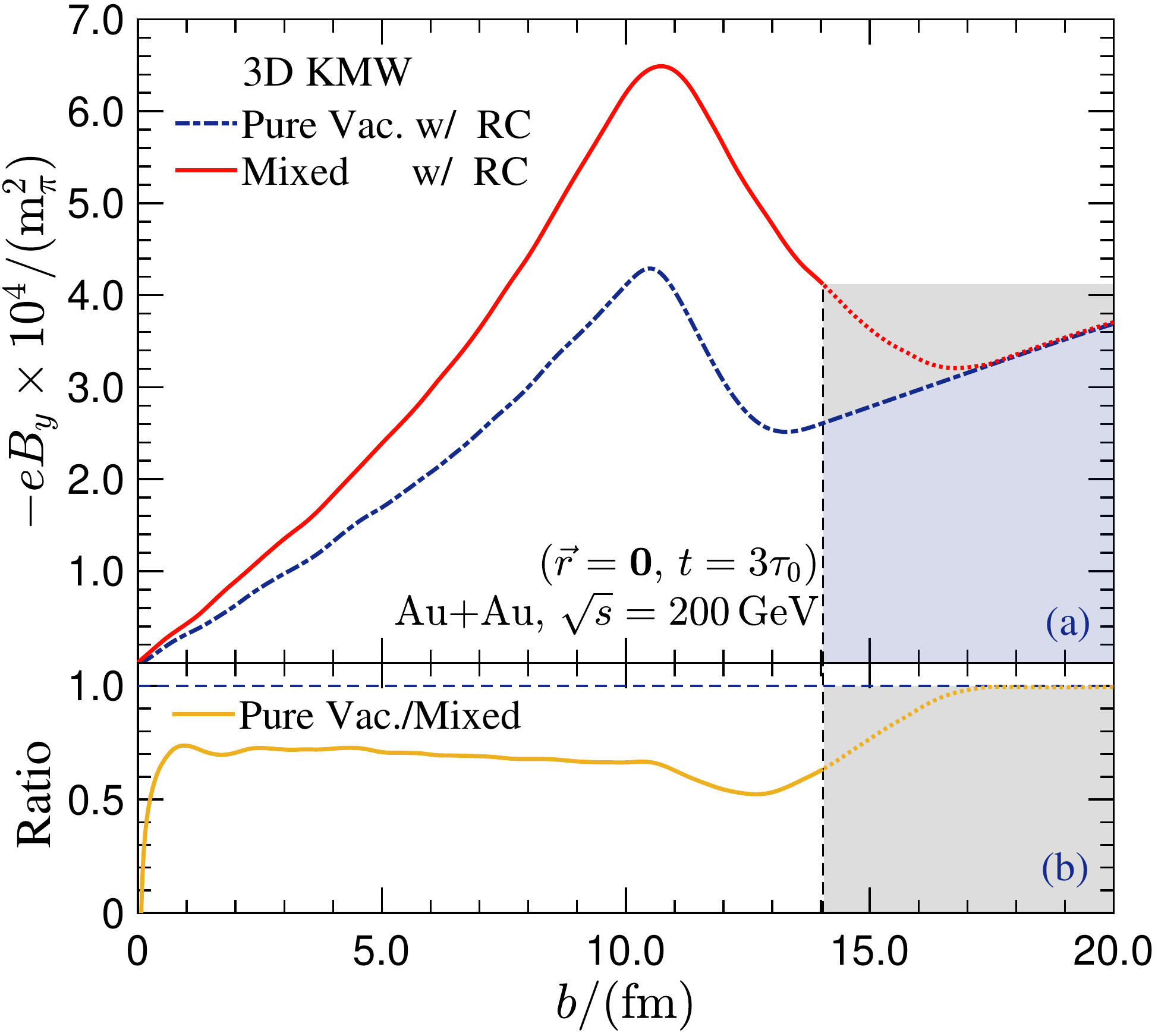}
        \caption{(Color online) (a) Comparison of impact parameter $b$ dependence of the total magnetic field strength with retardation correction at $t=3\tau_{0}$~\cite{HChao2008,Peter2003} from the extended (3D) KMW model during the realistic QGP evolution (denoted as ``Mixed") with that in the pure vacuum (denoted as ``Pure Vac.") at the center point $\mathbf{r}=\mathbf{0}$ of the overlapping region in Au+Au collisions at $\sqrt{s}=200\mathrm{\,GeV}$. (b) The corresponding ratio (Pure Vac./Mixed) of the total magnetic field strength with (w/) retardation correction (RC) as a function of impact parameter $b$.}
        \label{Fig06_eBCenMixed}
        \end{figure}

        In Fig. \ref{Fig06_eBCenMixed}, we show the impact parameter $b$ dependence of the total magnetic field strength at $t=3\tau_{0}$~\cite{HChao2008,Peter2003} during the realistic QGP evolution with a time-dependent step-function-like electric conductivity $\tilde{\sigma}(t)=\sigma_{0}\cdot\theta(t-t_{\sigma})$ (denoted as ``Mixed") from the extended KMW model in Eqs. (\ref{MaxR_SP01}-\ref{MaxR_ST02}) with retardation correction, compared with that in the pure vacuum (denoted as ``Pure Vac.") in Eqs. (\ref{LS_BS05}-\ref{LS_PT03}). The observation space point is $\mathbf{r}=\mathbf{0}$ in Au+Au collisions at $\sqrt{s}=200\mathrm{\,GeV}$. One can clearly see that the total magnetic field strength in the mixed scenario is generally larger than that in the pure vacuum scenario, indicating that medium feedback effects have significantly enhanced the magnetic field strength at later-time stages, e.g. at $t=3\tau_{0}$.

        Here we should note that at $b>2R_{\mathrm{A}}$, there is no longer any contributions of participants since the two colliding nuclei indeed do not collide with each other at all according to the spirit of KMW model. Thus, the conducting QGP medium in principle will not be created, for which the pure vacuum result in Fig. \ref{Fig06_eBCenMixed} is still appropriate since the magnetic field in the pure vacuum scenario does not depend on whether the conducting QGP medium is created or not. However, it will be a little different for the realistic QGP evolution scenario because the magnetic field in this scenario will depend on the formation of the conducting QGP medium, which thus indicates that the mixed scenario approximation is not appropriate and suitable at $b>2R_{\mathrm{A}}$, as is indicated by the grey shaded band.

        \begin{figure}[htb]
        \includegraphics[angle=0,scale=0.48]{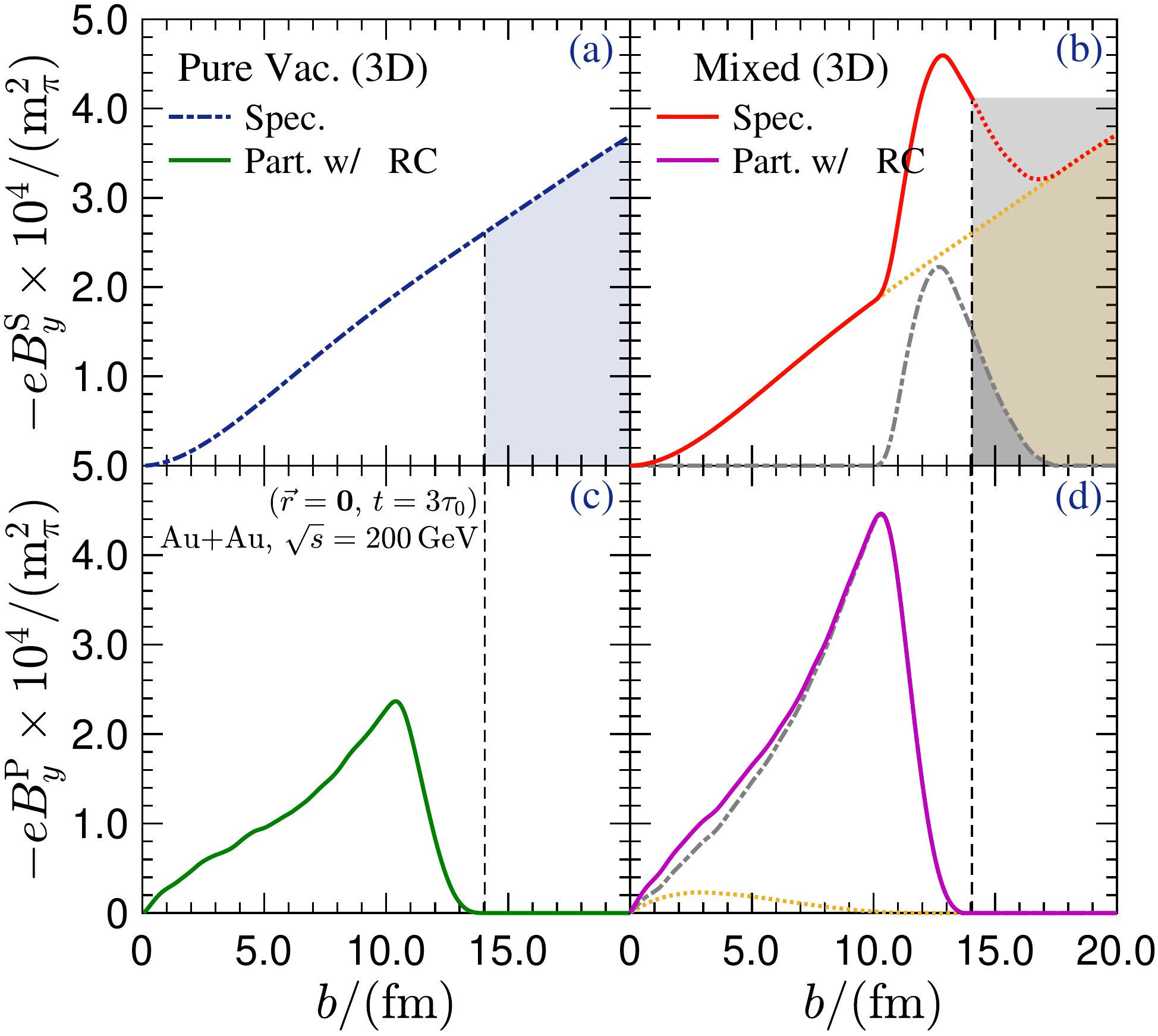}
        \caption{(Color online) Comparison of impact parameter $b$ dependence of the magnetic field strength contributed by spectators (S) in the upper panels (a) and (b), and that by participants (P) in the lower panels (c) and (d) at $t=3\tau_{0}$~\cite{HChao2008,Peter2003} from the extended (3D) KMW model during the realistic QGP evolution (denoted as ``Mixed") with that in the pure vacuum (denoted as ``Pure Vac.") at the center point $\mathbf{r}=\mathbf{0}$ of the overlapping region in Au+Au collisions at $\sqrt{s}=200\mathrm{\,GeV}$. The dotted yellow (dotted-dash gray) lines in the right panels represent the contributions of source charges of spectators or participants in the vacuum (conducting QGP medium).}
        \label{Fig07_eBCenMixedSP}
        \end{figure}

        Likewise, we show in Fig. \ref{Fig07_eBCenMixedSP} detailed comparisons of the impact parameter $b$ dependence of the estimated magnetic field strength with retardation correction at $t=3\tau_{0}$ contributed by spectators and participants during the realistic QGP evolution from the extended KMW model in Eqs. (\ref{MaxR_SP01}-\ref{MaxR_ST02}) (denoted as ``Mixed") with that in the pure vacuum (denoted as ``Pure Vac.") in Eqs. (\ref{LS_BS05}-\ref{LS_PT03}). The results in Fig. \ref{Fig07_eBCenMixedSP} are certainly the decomposition of the total magnetic field strength in Fig. \ref{Fig06_eBCenMixed}. In the upper panels of Fig. \ref{Fig07_eBCenMixedSP}, one can clearly see that the magnetic field strength from spectators in the pure vacuum scenario at $t= 3\tau_{0}$ is monotonically increasing with impact parameter $b$, while that in the mixed scenario is non-monotonically increasing with impact parameter $b$ especially at $b\sim13\mathrm{\,fm}$. We indeed check that in the mixed scenario, the contributions of magnetic field strength from source charges of spectators in the pure vacuum are also monotonically increasing with impact parameter $b$, see the dotted yellow line in Fig. \ref{Fig07_eBCenMixedSP} (b). It is due to the contributions of source charges of spectators in the conducting QGP medium receiving medium feedback effects that finally result in the non-monotonic peak around $b\sim13\mathrm{\,fm}$, see the dotted-dash grey line in Fig. \ref{Fig07_eBCenMixedSP} (b). Note that at $b>2R_{\mathrm{A}}$, the conducting QGP medium in principle will not be created according to the KMW model, thus the dotted yellow line for the pure vacuum scenario in Fig. \ref{Fig07_eBCenMixedSP} (b) is still valid while the dotted-dash grey line for the conducting QGP medium scenario approximation is no longer appropriate and suitable at $b>2R_{\mathrm{A}}$. Thus the estimated total contributions of spectators in the mixed scenario is not appropriate at $b>2R_{\mathrm{A}}$, as is indicated by the gray shaded band Fig. \ref{Fig07_eBCenMixedSP} (b). This is also the reason why the predicted total magnetic field strength in Fig. \ref{Fig06_eBCenMixed} for the mixed scenario is not appropriate.

        In the lower panels of Fig. \ref{Fig07_eBCenMixedSP}, we find the magnetic field strength contributed by participants at $t= 3\tau_{0}$ in the mixed scenario is significantly larger than that in the pure vacuum scenario, especially at $b\sim 10\mathrm{\,fm}$. Meanwhile, we notice that the total contribution of participants in the mixed scenario is dominated by source charges in the conducting medium, see the dotted-dash grey line in Fig. \ref{Fig07_eBCenMixedSP} (d). Thus, we realize that the medium feedback effects have become quite important at $t= 3\tau_{0}$. Here we should note that the observation time $t$ can also affect the observed centrality dependence of magnetic field, e.g. the shapes of impact parameter dependence of magnetic field strength in lower panels of Fig. \ref{Fig07_eBCenMixedSP} are rather sharp than that in Fig. \ref{Fig05_eBCenSP}. Such a phenomenon can be verified by taking different time slices during the time evolution so as to compare its impact parameter dependence. By doing so, one may find that the total magnetic field strength is roughly proportional to the impact parameter $b$ for smaller impact parameters, and such a dependence should also be sensitive to the observation time $t$ since the time dependence of magnetic field strength is quite significant in relativistic heavy-ion collisions.

\section{Summary}
\label{sec:Summary}

     In summary, based on the Kharzeev-McLerran-Warringa (KMW) model~\cite{Kharzv2008} for the estimation of strong EM fields generated in relativistic heavy-ion collisions, we make an attempt to generalize the formulas of estimated EM fields in the original KMW model, which eventually turns out to be above formulations in the extended KMW model consisting of three scenarios: the pure vacuum scenario, the pure conducting medium scenario and the relatively realistic QGP evolution scenario.

     We first start from generalizing the widely used charge distributions by incorporating the Lorentz contraction effect for the ellipsoidal geometry of colliding nuclei along with the deformation effect for heavy-ion collisions. We then combine the widely used L-W equations of EM fields for point-like charges with the generalized three-dimensional charge distributions, which can naturally generalize the formulas of EM fields in the vacuum by incorporating the longitudinal position dependence and retardation correction. Since the medium feedback effects for a conducting medium may substantially modify the time dependence of the generated EM fields according to the Faraday's induction law, we then explicitly and analytically embed a constant Ohm electric conductivity $\sigma_{0}$ into the Maxwell's equations for the incorporation of medium effects, and eventually formulate the estimation of EM fields in the extended KMW model for a pure conducting medium scenario with a constant Ohm electric conductivity $\sigma_{0}$, and also for a relatively realistic QGP evolution with a time-dependent electric conductivity $\tilde{\sigma}(t)$. For simplicity, we also propose one possible simplification of the formulas of EM fields during the realistic QGP evolution in APPENDIX A, which can serve as an alternative solution to doing Monte-Carlo simulations of EM filds instead of the L-W equations for heavy-ion collisions. Thus, we systematically formulate the estimations of EM fields in the extended KMW model.

     Our formulations in the extended KMW model can result in a slower time evolution (or longer lifetime $t_{\mathrm{B}}\sim 5\mathrm{\,fm/{\it c}}$) and a more reasonable impact parameter $b$ dependence of the magnetic field strength in the vacuum. The inclusion of medium effects from lattice QCD results on electric conductivity helps further prolong the time evolution of magnetic field. From numerical evaluations, we find that the pure vacuum approximation at early-time stages is a superior approximation, which is actually comparable with the mixed scenario approximation for a realistic QGP evolution. The pure conducting medium approximation, however, is not an appropriate approximation at early-time stages, which will overestimate the magnetic field strength by orders of magnitude. For later-time stages QGP evolution, the pure conducting medium approximation might be a good approximation while the pure vacuum approximation is no longer appropriate and suitable. Thus at thermal freeze-out time, the predicted magnetic field strength in the realistic QGP evolution scenario can meet the $1\sigma$ bound constrained from experimental measurements on global polarizations of $\Lambda$ and $\bar{\Lambda}$ hyperons in Au+Au collisions at top RHIC energy. However, since the magnetic field may not be the only possible explanation and there are some limitations in our model, we only expect at best our model might semi-quantitatively describe the polarization difference at top RHIC energy.

     It should be emphasized that our generalization of charge distributions with the three-parameter Fermi model that we present in this paper is actually a sample example, many other charge distribution models that are listed in~\cite{DVries1987} or somewhere else for various colliding nuclei can be similarly generalized in the same way as we present here. Last but not least, the extended KMW model can be potentially applied to various colliding energies, especially for the lower energy regions, like the ongoing STAR-BES program and under planning FAIR, NICA and J-PARC programs. Hopefully, these generalized formulations in the extended KMW model can be employed for many EM fields relevant studies, such as the CME related charge asymmetry correlators or fluctuations in the experimental searching, modifications of in-medium particle's mass as well as the QCD phase diagram under strong magnetic field, and so on.

\begin{acknowledgements}

     The authors would like to gratefully thank Yu-Gang Ma, Jin-Hui Chen, Song Zhang and Qi-Ye Shou for their stimulating and helpful discussions and comments, and Chen Zhong for maintaining high-quality performance of computational facilities. Y. C. thankfully acknowledges the helpful and fruitful discussions with Wei-Tian Deng, Huan-Xiong Yang, Heng-Tong Ding, Xu-Guang Huang, and Qun Wang, as well as the timely discussions and helps from Xin-Li Zhao, Yi-Lin Cheng, Xin Ai, Bang-Xiang Chen, Yu Guo, Xiao-Liang Xia, Hui Li and Xian-Gai Deng.  Y. C. and G.-L. M. are supported by National Natural Science Foundation of China (NSFC) under Grants Nos. 11890710, 11890714, 11835002, and 11961131011, the Strategic Priority Research Program of Chinese Academy of Sciences under Grant No. XDB34030000, and the Guangdong Major Project of Basic and Applied Basic Research under Grant No. 2020B0301030008. X.-L. S. is supported by National Natural Science Foundation of China (NSFC) under Grant Nos. 11890714 and 12047528.

\end{acknowledgements}
~\\

\appendices
\makeatletter 
\@addtoreset{equation}{section}
\makeatother  
\renewcommand\thesection{APPENDIX~\Alph{section}}
\renewcommand\theequation{\Alph{section}.\arabic{equation}}

\section{An Alternative Solution to the Estimated EM Fields in Simulations}
\label{sec:An Alternative Solution to the Estimated EM Fields in Simulations}

      Due to the ellipsoidal-type geometry and also peculiar charge domains occupied by spectators and participants, the numerical integrations of charge distribution for the estimated EM fields in the pure vacuum in Eqs. (\ref{LS_BS05}-\ref{LS_PT03}), or in the pure conducting medium in Eqs. (\ref{Max_SP01}-\ref{Max_ST02}), or during the realistic QGP evolution in Eqs. (\ref{MaxR_SP01}-\ref{MaxR_ST02}) are actually not very easily and quickly performed. Meanwhile, we notice that the L-W equations of EM fields in the vacuum, i.e. Eq. (\ref{LS_BS01}), are so widely accepted and used~\cite{Skokov2009,Sergei2010,LiOou2011,Bzdak2012,DenHu2012,ToneA2012,Blocz2013,DenHu2015,Blocz2015,VecRoy2015,XLZhao2018,XLZhaB2019,XLZhaC2019,Hamelm2019,YChen2019,XianG2020}. We therefore propose one possible simplification of the formulas of EM fields during the realistic QGP evolution from Eqs. (\ref{MaxR_SP01}-\ref{MaxR_ST02}) by replacing the continuous integration of charge distribution with the discrete summation for all point-like charges, which turns out to be an alternative solution to doing Monte-Carlo numerical simulations of generated EM fields during the realistic QGP evolution instead of the L-W equations as follows

      \begin{equation}\label{Appex_LW01}
      \begin{aligned}
      e\mathbf{B}_{\mathrm{S}}^{\pm}(t, \mathbf{r}) &=  \lim_{Y\to \pm Y_0} \alpha_{\mathrm{EM}} \sum_{N_{\mathrm{S}}} Z_n \frac{ \sinh(Y) \cdot \mathbf{e}_{z} \times \mathbf{R}_{\pm} }{ \Delta^{3/2} }\left[1 + \frac{\tilde{\sigma}(t_{\mathrm{ret}})\sinh|Y| }{2}\sqrt{\Delta} \right]e^{\tilde{A}},\\
      e\mathbf{B}_{\mathrm{P}}^{\pm}(t, \mathbf{r}) &=  \alpha_{\mathrm{EM}}\sum_{N_{\mathrm{P}}} Z_n \int_{-Y_0}^{Y_0} \mathrm{d}Y\,\frac{ \Psi_{\pm}(Y)\sinh Y \cdot \mathbf{e}_{z} \times \mathbf{R}_{\pm} }{ \Delta^{3/2} }\left[1 + \frac{\tilde{\sigma}(t_{\mathrm{ret}})\sinh|Y| }{2}\sqrt{\Delta}\right]e^{\tilde{A}},
      \end{aligned}
      \end{equation}
      for the decomposed magnetic field $\bf{B}(t, \mathbf{r})$ contributed by spectators and participants. For that of the electric field $\mathbf{E}(t, \mathbf{r})$, we obtain the following expressions for $x$ and $y$ components of the electric field $E_{x}$ and $E_{y}$,
      \begin{equation}\label{Appex_LW02}
      \begin{aligned}
      \begin{pmatrix} e{E}_{x,\mathrm{S}}^{\pm}\\ e{E}_{y,\mathrm{S}}^{\pm} \end{pmatrix}(t, \mathbf{r}) &= \lim_{Y\to \pm Y_0} \alpha_{\mathrm{EM}} \sum_{N_{\mathrm{S}}} Z_n
      \bigg{\{}\frac{ R_{\perp}\cosh Y}{\Delta^{3/2}}\left(1 + \frac{\tilde{\sigma}(t_{\mathrm{ret}})\sinh |Y|}{2}\sqrt{\Delta}\right)\\
      \quad\quad\quad&\quad - \frac{\tilde{\sigma}(t_{\mathrm{ret}})}{R_{\perp} \tanh |Y|} \left[1 + \frac{\sinh | Y|}{\sqrt{\Delta}} \left(t - \frac{z - z^{\prime}}{\tanh Y}\right) \right] \exp\left(-\tilde{\sigma}(t_{\mathrm{ret}})\left[ t - \frac{z-z^{\prime}}{\tanh Y} \right]\right) \bigg{\}}\frac{e^{\tilde{A}}}{R_{\perp}}\begin{pmatrix} x-x^{\prime}\\ y-y^{\prime} \end{pmatrix},\\
      \begin{pmatrix} e{E}_{x,\mathrm{P}}^{\pm}\\ e{E}_{y,\mathrm{P}}^{\pm} \end{pmatrix}(t, \mathbf{r}) &= \alpha_{\mathrm{EM}} \sum_{N_{\mathrm{P}}} Z_n \int_{-Y_0}^{Y_0} \mathrm{d}Y\,\Psi_{\pm}(Y)\bigg{\{}\frac{ R_{\perp}\cosh Y}{\Delta^{3/2}}\left(1 + \frac{\tilde{\sigma}(t_{\mathrm{ret}})\sinh |Y|}{2}\sqrt{\Delta}\right)\\
      \quad\quad\quad&\quad - \frac{\tilde{\sigma}(t_{\mathrm{ret}})}{R_{\perp} \tanh |Y|} \left[1 + \frac{\sinh |Y|}{\sqrt{\Delta}} \left(t-\frac{z - z^{\prime}}{\tanh Y}\right) \right] \exp\left(-\tilde{\sigma}(t_{\mathrm{ret}})\left[ t - \frac{z-z^{\prime}}{\tanh Y} \right]\right)\bigg{\}}\frac{e^{\tilde{A}}}{R_{\perp}}\begin{pmatrix} x-x^{\prime}\\ y-y^{\prime} \end{pmatrix},
      \end{aligned}
      \end{equation}
      and also that for the $z$ component of the electric field $E_{z}$
      \begin{equation}\label{Appex_LW03}
      \begin{aligned}
      e{E}_{z,\mathrm{S}}^{\pm}(t, \mathbf{r}) &= \lim_{Y\to \pm Y_0} \alpha_{\mathrm{EM}} \sum_{N_{\mathrm{S}}} Z_n
      \bigg{\{} \frac{\mathrm{sgn}(Y) \tilde{\sigma}^2(t_{\mathrm{ret}})}{\tanh^2 Y} \exp\left(-\tilde{\sigma}(t_{\mathrm{ret}})\left[ t - \frac{z-z^{\prime}}{\tanh Y} \right]\right) \Gamma(0, -\tilde{A})\\
      \quad\quad\quad&\quad + \frac{e^{\tilde{A}}}{{\Delta}^{3/2}}\left[(z-z^{\prime})\cosh Y - t\sinh Y - \mathrm{sgn}(Y)\tilde{A} \sqrt{\Delta} - \frac{\tilde{\sigma}(t_{\mathrm{ret}})\sinh Y}{\tanh^2 Y}\Delta \right]  \bigg{\}},\\
      e{E}_{z,\mathrm{P}}^{\pm}(t, \mathbf{r}) &= \alpha_{\mathrm{EM}} \sum_{N_{\mathrm{P}}} Z_n \int_{-Y_0}^{Y_0} \mathrm{d}Y\, \Psi_{\pm}(Y)\bigg{\{}
      \frac{\mathrm{sgn}(Y)\tilde{\sigma}^2(t_{\mathrm{ret}})}{\tanh^2 Y} \exp\left(-\tilde{\sigma}(t_{\mathrm{ret}})\left[ t - \frac{z-z^{\prime}}{\tanh Y} \right]\right) \Gamma(0, -\tilde{A})\\
      \quad\quad\quad&\quad +\frac{e^{\tilde{A}}}{{\Delta}^{3/2}} \left[ (z-z^{\prime})\cosh Y - t\sinh Y - \mathrm{sgn}(Y) \tilde{A} \sqrt{\Delta} - \frac{\tilde{\sigma}(t_{\mathrm{ret}})\sinh Y}{\tanh^2 Y}\Delta \right]  \bigg{\}}.
      \end{aligned}
      \end{equation}
      Here the expressions for $R_{\perp}$, $\Delta$ and $\tilde{A}$ are the same as the definitions in Eq. (\ref{MaxR_SP02}), and $\Psi_{\pm}(Y)$ in Eq. (\ref{LS_PT03}). $N_{\mathrm{S}}$ and $N_{\mathrm{P}}$ denote the total numbers of spectators and participants of point-like charges, respectively. Also, we note that a similar Monte-Carlo simulation using above formulas of EM fields for point-like charges in Eqs. (\ref{Appex_LW01}-\ref{Appex_LW03}) has firstly been performed in~\cite{LiHui2016}, since such an alternative formulation could have been more directly obtained from the EM fields of point-like charges in~\cite{LiHui2016} when further embedded with the refined baryon-junction stopping effect with retardation correction as in Eq. (\ref{LS_PT03}) and time-dependent electric conductivity $\tilde{\sigma}(t) = \sigma_{0}\cdot\theta(t-t_{\sigma})$.

      If one assumes that each proton can be treated as point-like charge, as has been extensively assumed in literatures~\cite{Skokov2009,Sergei2010,LiOou2011,Bzdak2012,DenHu2012,ToneA2012,Blocz2013,DenHu2015,VecRoy2015,XLZhao2018,XLZhaB2019,XLZhaC2019,Hamelm2019,YChen2019,XianG2020}, then above generalization of EM fields in Eqs. (\ref{Appex_LW01}-\ref{Appex_LW03}) for Monte-Carlo simulations in heavy-ion collisions can have at least two merits: one is that the EM fields from Eqs. (\ref{Appex_LW01}-\ref{Appex_LW03}) are analytically embedded with a step-function-like electric conductivity $\tilde{\sigma}(t)$ for the incorporation of medium effects during the QGP evolution rather than the original L-W equations in the vacuum; the other is that the EM fields in Eqs. (\ref{Appex_LW01}-\ref{Appex_LW03}) are also implanted with the refined baryon-junction stopping effect with retardation correction as in Eq. (\ref{LS_PT03}).


\end{document}